\documentclass[aip,rsi,amsmath,amssymb,reprint]{revtex4-2}

\usepackage{graphicx}
\usepackage{dcolumn}
\usepackage{bm}

\usepackage[utf8]{inputenc}
\usepackage[T1]{fontenc}
\usepackage{mathptmx}
\usepackage{etoolbox}
\usepackage{booktabs} 
\usepackage{tabularx,makecell,multirow} 

\newcommand{\sub}{\textsubscript}

\begin{document}


\title[A Cryogenically-Cooled High-Sensitivity Nuclear Quadrupole Resonance Spectrometer]{A Cryogenically-Cooled High-Sensitivity Nuclear Quadrupole Resonance Spectrometer} 



\author{Jarred Glickstein}%
\email{jsg109@case.edu}
\author{Soumyajit Mandal}%
\altaffiliation[Author ]{to whom correspondence should be addressed}
\email{soumya@alum.mit.edu}
\affiliation{Electrical, Computer, and Systems Engineering Department, Case Western Reserve University, Cleveland, OH 44106, USA.}


\date{\today}

\begin{abstract}
The paper describes a radio frequency (RF) spectrometer for $^{14}$N nuclear quadrupole resonance (NQR) spectroscopy that uses a detector coil cooled to 77~K to maximize measurement sensitivity. The design uses a minimally-intrusive network of active duplexers and mechanical contact switches to realize a digitally reconfigurable series/parallel coil tuning network that allows transmit- and receive-mode performance to be independently optimized. The design is battery-powered and includes a mixed-signal embedded system to monitor and control secondary processes, thus enabling autonomous operation. Tests on an acetaminophen sample show that cooling both the detector and sample increases the signal-to-noise ratio (SNR) per scan by a factor of approximately 88 (in power units), in good agreement with theoretical predictions.
\end{abstract}

\pacs{}

\maketitle 

\section{Introduction}
\label{sec:intro}

Nuclear quadrupole resonance (NQR) spectroscopy is a quantitative and non-destructive radio frequency (RF) analytical technique with applications in solid-state chemistry~\cite{weiss1990correlation,szell2016solid}, detection of explosives~\cite{gudmundson2009nqr}, characterization of pharmaceuticals~\cite{chen2015authentication}, and authentication of manufactured objects~\cite{masna2019robust,masna2021nqr}. While a variety of quadrupolar nuclei generate NQR spectra, $^{14}N$ is often of primary interest due to its widespread occurrence in organic compounds, pharmaceuticals, and explosives. However, the applications of $^{14}N$ NQR spectroscopy are often limited by its intrinsically low sensitivity compared to both optical methods (e.g., near-infrared and Raman spectroscopy) and nuclear magnetic resonance (NMR) spectroscopy. The key issue is that the separations between nuclear energy levels involved in NQR transitions are small (in the low-MHz range for $^{14}N$), which results in small population differences in thermal equilibrium and thus poor signal-to-noise ratios (SNR) for the transitions. Moreover, these separations are intrinsic to the sample and its crystal structure and cannot be increased by applying an external field, unlike for NMR.

The low sensitivity of $^{14}N$ NQR spectroscopy makes it particularly time-consuming to discover the spectra of hitherto-unknown samples, since this generally requires broadband high-resolution rfrequency sweeps. Reducing the temperature of the sample and/or detector is an effective way to decrease overall experimental time by increasing the SNR per scan. Accordingly, laboratory applications of NQR spectroscopy have long relied on cryostats to cool the sample~\cite{albert1978design}. In~\cite{ostafin1994automatically}, the authors describe a conventional tuned (i.e., narrowband) inductive detector within a cryostat; the tuning frequency can be programmed via by a mechanically-variable capacitor controlled by a stepper motor. Other work has focused on broadband superconducting detectors (typically operating at 4~K), such as SQUIDs~\cite{miller2007nuclear} and superconducting-magneto-resistive hybrid sensors~\cite{pannetier200914}. However, mechanical tuning methods are slow ($\sim 10-100$~ms per frequency step), thus making it difficult to use frequency-interleaved scans~\cite{nqr:cheng2017_vitamins} to speed up frequency sweeps. On the other hand, broadband detectors generally require extensive magnetic shielding to protect against external radio frequency interference (RFI).

In this paper, we address these issues by developing a high-sensitivity $^{14}N$ NQR spectrometer that uses a custom cryostat to cool both the sample and an inductive detector (coil) to liquid-$N_{2}$ temperatures (77~K). In addition, a flexible series/parallel electronic tuning network enables rapid probe re-tuning and allows the transmit- and receive-mode performance of the coil to be independently optimized. The rest of the paper is organized as follows. Section~\ref{sec:theory} discusses the sensitivity of NQR spectroscopy and its dependence on temperature. The proposed tunable probe is discussed in Section~\ref{sec:probe_design}. Section~\ref{sec:nqr_receiver_2} and \ref{sec:nqr_transmitter_v2} discuss the custom RF receiver and transmitter used by the proposed spectrometer. System-level implementation and test results are described in Section~\ref{sec:nqr_gen2_spectrometer}, while Section~\ref{sec:conclusion} summarizes our conclusions.

\section{Theoretical Analysis}
\label{sec:theory}

\subsection{Sensitivity of NQR Spectroscopy}
The sensitivity of NQR spectroscopy can be theoretically estimated as follows. We begin by estimating the magnetization density within the sample immediately after an RF excitation pulse with a nutation angle of $\theta$. Assuming that i) the sample was originally in thermal equilibrium, and ii) the RF pulse has enough bandwidth to excite the entire width of an NQR line centered around $\omega_0$, the result is~\cite{dehmelt1954nuclear,nqr:cheng2017_vitamins}
\begin{equation}
    M_{0}(\theta)=s_{exc}(\theta)\rho_s\frac{\gamma\hbar^2\omega_0}{(2I+1)k_{B}T_s}\left(\frac{I(I+1)-mm^{\prime}}{I^2}\right),
    \label{eq:m0}
\end{equation}
where $s_{exc}(\theta)$ is known as the nutation function, $\rho_s$ is the spin density (i.e., the number of target nuclei per unit volume), $\gamma$ is the gyromagnetic ratio of the nucleus, $\hbar$ is the reduced Planck's constant, $k_{B}$ is Boltzmann's constant, $T_s$ is the sample temperature, $I\geq 1$ is the total angular momentum (i.e., nuclear spin), and $m$ and $m^{\prime}$ are the spin quantum numbers of the two nuclear states involved in the NQR transition. Due to quantum selection rules, these values must satisfy $|m-m^{\prime}|=1$.  For spin-1 nuclei such as $^{14}$N (i.e., nuclei with $I=1$), the frequencies of the three allowed transitions are conventionally denoted by $\nu_+$, $\nu_-$, and $\nu_0$, respectively.

For polycrystalline materials, the nutation function $s_{exc}(\theta)$ must be calculated via a so-called powder average over all possible relative orientations of the crystallographic principal axes to the RF magnetic field generated by the pulse, $\vec{B}_{1}$. For spin-1 nuclei, the result can be shown to be~\cite{lee2002spin}
\begin{equation}
    s_{exc}(\theta)\approx\sqrt{\frac{\pi}{2\theta}}J_{3/2}(\theta),
\end{equation}
where $J_{3/2}(\theta)$ denotes the Bessel function of the first kind. The peak value of this function is $\approx 0.426$, which occurs at the optimum nutation angle of $\theta_{opt}\approx 2.09$~rad (i.e., $119.5^{\circ}$).

The induced magnetization, $M_0$, represents the initial amplitude of a quantum coherence that oscillates sinusoidally at the resonant frequency, $\omega_0$. Here we assume that it is measured using a conventional inductive detector (i.e., coil). By Faraday induction, the resulting time-varying magnetic flux generates a RF voltage in the detector coil. Using the principle of reciprocity for electromagnetic fields~\cite{hoult1976signal}, the amplitude of this voltage is given by
\begin{equation}
    V_{coil}=\omega_0\int_{V_s}\left(\frac{B_1(\vec{r})}{I_1}\right)M_0(\theta(\vec{r}))dV_{s},
    \label{eq:vcoil}
\end{equation}
where the integral is carried out over $V_{s}$, the sample volume; $\theta(\vec{r})=\sqrt{3}B_{1}(\vec{r})t_p$ is the position-dependent nutation angle (with $t_p$ being the RF pulse length); and $B_{1}(\vec{r})/I_{1}$ is the position-dependent \emph{coil sensitivity function}, i.e., the amplitude of the RF field generated by a unit current flowing in the coil.

Consider a pulsed NQR experiment with spin echo detection, often known as a spin-locked spin echo (SLSE) or pulsed spin locking (PSL) pulse sequence~\cite{fraissard2009explosives}. Ignoring transverse relaxation, the signal-to-noise ratio (SNR) per echo (in power units) is then given by
\begin{equation}
    SNR_{e} = \frac{V_{coil}^{2}}{2\sigma^{2}_{n}},
\end{equation}
where $\sigma_{n}$ is the root-mean-squared (rms) noise in the coil and the factor of 2 arises from converting amplitude to rms. In the absence of external sources (such as RFI), $\sigma_{n}$ is dominated by the thermal noise of the coil and given by
\begin{equation}
    \sigma^{2}_{n}=4k_{B}T_{c}R_{c}F\Delta f,
    \label{eq:noise}
\end{equation}
where $T_c$ is the coil temperature, $R_c$ is the series coil resistance, $F>1$ is the noise factor of the receiver, and $\Delta f$ is the final detection bandwidth. Note that the coil resistance can be written as $R_{c}=\omega_0 L_c/Q_c$, where $L_c$ and $Q_c$ are the coil inductance and quality factor at $\omega_0$, respectively. Also $\Delta f\approx \left(1/T_{2}^{\ast}+1/T_{acq}\right)$, where $T_{2}^{\ast}$ is the decay constant due to inhomogeneous broadening of the NQR line (which determines the width of each spin echo) and $T_{acq}$ is the duration of the echo acquisition window.

In practice, the SNR obtained after averaging across several scans is reduced by i) signal decay due to transverse relaxation during each SLSE sequence (defined by the time constant $T_{2,eff}$), and ii) wait times $t_W$ between SLSE sequences to allow for longitudinal ($T_{1}$) relaxation. This effect has been analyzed in detail in our earlier work~\cite{masna2019robust}; the SNR available per echo after averaging is found to be
\begin{equation}
    SNR_{e,av} = SNR_{e}\frac{\left(1-e^{-\alpha}\right)^2\left(1-e^{-\beta}\right)^2}{\beta\left(\beta+\alpha\delta\right)},
    \label{eq:snr_av}
\end{equation}
where $\alpha=t_{W}/T_{1}$, $\beta=t_{SLSE}/T_{2,eff}$, and $\delta=T_{1}/T_{2,eff}$ are dimensionless parameters, and $t_{SLSE}=N_{E}t_{E}$ is the total duration of a single SLSE sequence (consisting of $N_{E}$ echo periods, each of length $T_{E}$).

Given the measured relaxation time constants of the sample, the expression in eqn.~(\ref{eq:snr_av}) can be numerically maximized to find the measurement parameters that maximize $SNR_{e,av}$, i.e., to find the optimum values of $t_W$ (which controls $\alpha$) and $t_{SLSE}$ (which controls $\beta$)~\footnote{This process is complicated by the fact that $T_{2,eff}$ is a combination of several relaxation time constants, including $T_{1}$, $T_{2}$, $T_{1\rho}$ ($T_{1}$ in the rotating frame, which itself depends on SLSE pulse sequence parameters such as the $t_{E}$ and the peak RF power level), and $T_{2^{\ast}}$ (which depends on sample preparation). We will ignore the dependence of $T_{2,eff}$ on pulse sequence parameters for convenience; this is an acceptable approximation over commonly-used ranges of parameter values.}. For example, the $\nu_{+}$ transition of acetaminophen (in the monoclinic form) has $T_{1}\approx 9$~s and $T_{2,eff}\approx 2.5$~s for $t_{E}=1200$~$\mu$s at room temperature~\cite{luznik2013influence,masna2019robust}. The optimum parameter values are then found to be $\alpha_{opt}\approx 1.53$ and $\beta_{opt}=1.01$, resulting in $SNR_{e,av}\approx 0.038\times SNR_{e}$. 

\subsection{Effects of Temperature on Sensitivity}
\label{subsec:temp_sens}
Eqn.~(\ref{eq:m0}) shows that the magnetization density is inversely proportional to sample temperature, i.e., $M_{0}\propto 1/T_s$. Note that this relationship is ultimately derived from the relative populations of nuclear spin states in thermal equilibrium: $M_{0} \propto (n-n^{\prime})/n$ where $n$ and $n^{\prime}$ are the populations of the lower and upper energy states involved in the NQR transition. Boltzmann statistics predicts that $n^{\prime}/n=\exp(-\hbar\omega_{0}/k_BT_s)$, such that $M_{0}$ is proportional to $1 - \exp(-\hbar\omega_{0}/k_BT_s)\approx \hbar\omega_{0}/(k_BT_s)$. The latter approximation is valid when the energy difference between the states is much smaller than the thermal energy, i.e., $\hbar\omega_0 \ll k_BT_s$, which is always true for NQR transitions (with $\omega_0$ in the low-MHz range) for sample temperatures greater than a few mK.

Since the nutation function and coil sensitivity function are both independent of temperature, the induced coil voltage $V_{coil}$ is proportional to $M_0$. Thus, it also decreases with temperature as $V_{coil}\propto 1/T_s$. As result, cooling the sample increases the NQR signal power (and thus $SNR_e$) $\propto 1/T_s^{2}$.

Cooling the detector can further increase $SNR_e$ by reducing the noise power $\sigma^{2}_{n}$. Since the power spectral density (PSD) of coil thermal noise $\propto k_{B}T_{c}$, eqn.~(\ref{eq:noise}) predicts that $\sigma^{2}_{n}\propto T_c$ such that $SNR_e \propto 1/(T_s^2T_c)$. However, in reality the dependence is more complex since both the coil resistance $R_c$ and the receiver noise figure $F$ are temperature-dependent.

\begin{figure}
\centering
\includegraphics[width=0.85\columnwidth]{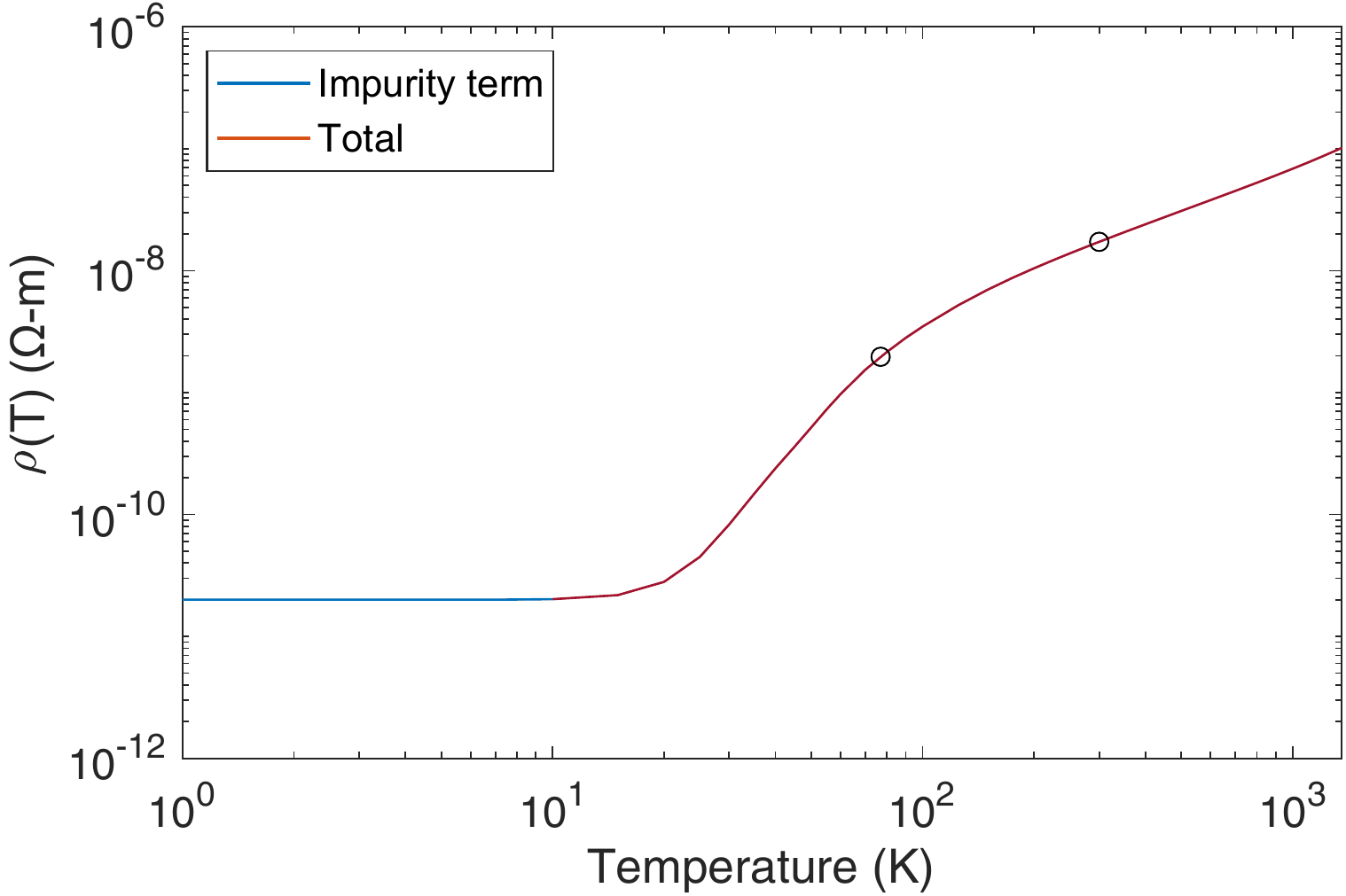}
\caption{Typical resistivity of annealed copper wire as a function of temperature, with the intrinsic resistivity $\rho_0 =2\times 10^{-11}$~$\Omega$-m. Data from~\cite{matula-resistivity-data}. The open circles denote room temperature (300~K) and liquid-$N_2$ temperature (77~K).}
\label{fig:resistivity_of_copper}
\end{figure}

First we analyze the coil resistance $R_c$, which is proportional to the resistivity of the coil winding material (typically copper). The resistivity $\rho$ of metals such as copper is dependent on the presence of impurities, which result in scattering of the carriers~\cite{matula-resistivity-data}. The contribution of such impurities to resistivity is approximately temperature-independent, such that the total resistivity versus temperature can be written as
\begin{equation}
    \rho(c,T) = \rho_0(c) + \rho_{i}(T),
\end{equation}
where $c$ is the impurity concentration and $\rho_0$ is the residual resistivity at 0K due to the impurities. For example, good-quality annealed copper wire has $\rho_0\approx 2\times 10^{-11}$~$\Omega$-m. In addition, $\rho_{i}$ is the intrinsic resistivity, which goes to zero at 0K approximately $\propto T^{5}$ and becomes $\propto T$ at high temperatures, i.e., as $T\rightarrow\infty$~\cite{matula-resistivity-data}. The resulting total resistivity $\rho(c,T)$ for annealed copper wire is shown in Fig.~\ref{fig:resistivity_of_copper}; it is approximately proportional to temperature (i.e., $\rho(T) \propto T$) above $\sim55$~K.

\begin{figure}
\centering
\includegraphics[width=0.8\columnwidth]{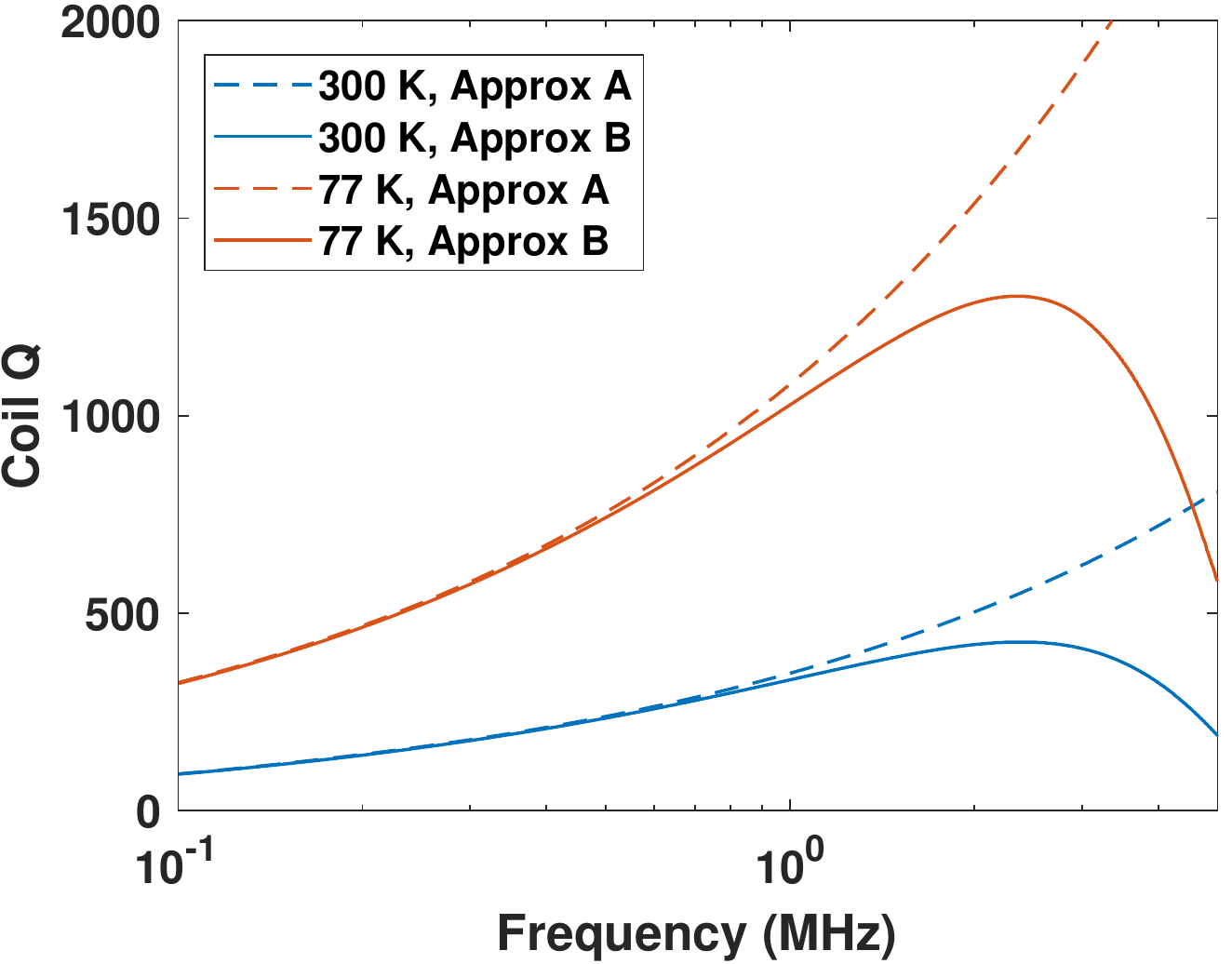}
\caption{Evaluation of two numerical models for a typical sample coil $Q$ at 77~K and 300~K. The two models treat the coil as i) an ideal inductor with series resistance $R_c$ (\textsf{Approx. A}), and ii) an ideal inductor with series $R_c$ and parallel capacitance $C_p$ (\textsf{Approx. B}).}
\label{fig:inductor_Q_numerical_model}
\end{figure}

The observed behavior of $\rho$ was included in a numerical model (referred to as \textsf{Approx. A}) that predicts the inductance and series resistance $R_c$ of typical solenoid sample coils as a function of their temperature $T_c$. The results can also be used to predict the coil Q, but the latter estimate becomes inaccurate at high frequencies due to the parasitic self-capacitance $C_p$ of the coil. The self-capacitance consists of both inter-turn capacitance, which can be reliably modeled~\cite{coildesign:stroobandt_hamwaves_calculator}, and poorly-modeled capacitive coupling to the shielding enclosure and other nearby conductors. Including an estimate of this term in the numerical model (referred to as \textsf{Approx. B}) provides a more accurate estimate of coil~Q versus frequency, as shown in Fig.~\ref{fig:inductor_Q_numerical_model} for a typical coil geometry at 77~K and 300~K. The model shows that cooling the coil using liquid-$N_2$, such that $T_{c}=77$~K, reduces $R_c$ (and thus increases $Q$) by approximately $3\times$ compared to room temperature. 

Next, we consider the receiver noise factor $F$. Defining the input-referred voltage and current noise PSDs of the receiver as $e_n^2$ and $i_n^2$, respectively, $F$ may be written as
\begin{equation}
    F = \left(1+\frac{e_n^2+i_n^2\left|Z_p\right|^2}{4k_{B}T_cR_cG_{p}^2}\right),
    \label{eq:rx_noise}
\end{equation}
where we have assumed that $e_n$ and $i_n$ (also known as the series and parallel noise, respectively) are uncorrelated for simplicity. In addition, $Z_p$ and $G_p$ are the output impedance and voltage gain of the NQR probe (which consists of the detector coil and an impedance matching or transformation network) around the resonant frequency $\omega_0$. Note that the coil thermal noise (i.e., the denominator in eqn.~(\ref{eq:rx_noise})) decreases strongly as the detector is cooled, making it more challenging to maintain a low receiver noise figure ($NF\equiv 10\log_{10}(F))$. In this case, one option is to reduce $e_n$ and $i_n$ by also cooling the receiver front-end electronics~\cite{cryocmos:proctor2015,pullia_cryogenic_jfet_preamplifier}. 

Here we consider the performance of the following common probe designs: i) un-tuned or broadband, with passive voltage gain provided by a transformer (turns ratio $1:N$); ii) tuned to $\omega_0$ with a parallel capacitor; and ii) matched to a reference impedance $Z_0$ (typically 50~$\Omega$) with a two-capacitor network. These cases, which have been analyzed in detail in~\cite{greer2021analytical}, result in the values of $Z_p$ and $G_p$ summarized in Table~\ref{tab:probe_properties}; here $L_c$ is the coil inductance.

\begin{table}[]
    \centering
    \caption{\label{tab:probe_properties}Properties of common NQR probe designs}
    \begin{tabular}{c|c|c}
    \hline
    \textbf{Probe topology} & \textbf{Voltage gain} & \textbf{Output impedance}\\
    \hline
    Un-tuned & $N$ & $j\omega_0 L_c N^2$\\
    Tuned & $Q$ & $\omega_0 L_c Q$\\
    Matched & $\frac{1}{2}\sqrt{Z_0/R_c}$ & $Z_0$\\
    \hline
    \end{tabular}
\end{table}

Table~\ref{tab:probe_properties} shows that tuned probes tend to have the largest voltage gain: typical values of $N$ (gain for un-tuned probes) and $\sqrt{Z_0/R_c}$ (gain for matched probes) range from 3-6 and 4-10, respectively, while $Q$ (gain for tuned probes) can reach values $>200$ at room temperature (see Fig.~\ref{fig:inductor_Q_numerical_model}). Eqn.~(\ref{eq:rx_noise}) then predicts that tuned probes have the lowest receiver NF. Equivalently, the high voltage gain of tuned probes enables their output impedance $\omega_0 L_c Q$ to approach the (usually) high noise resistance ($R_{n}\equiv e_n/i_n$) of FET-input receivers; as is well-known, satisfying this ``noise matching'' condition minimizes NF.

\begin{figure}
    \centering
    \includegraphics[width=0.75\columnwidth]{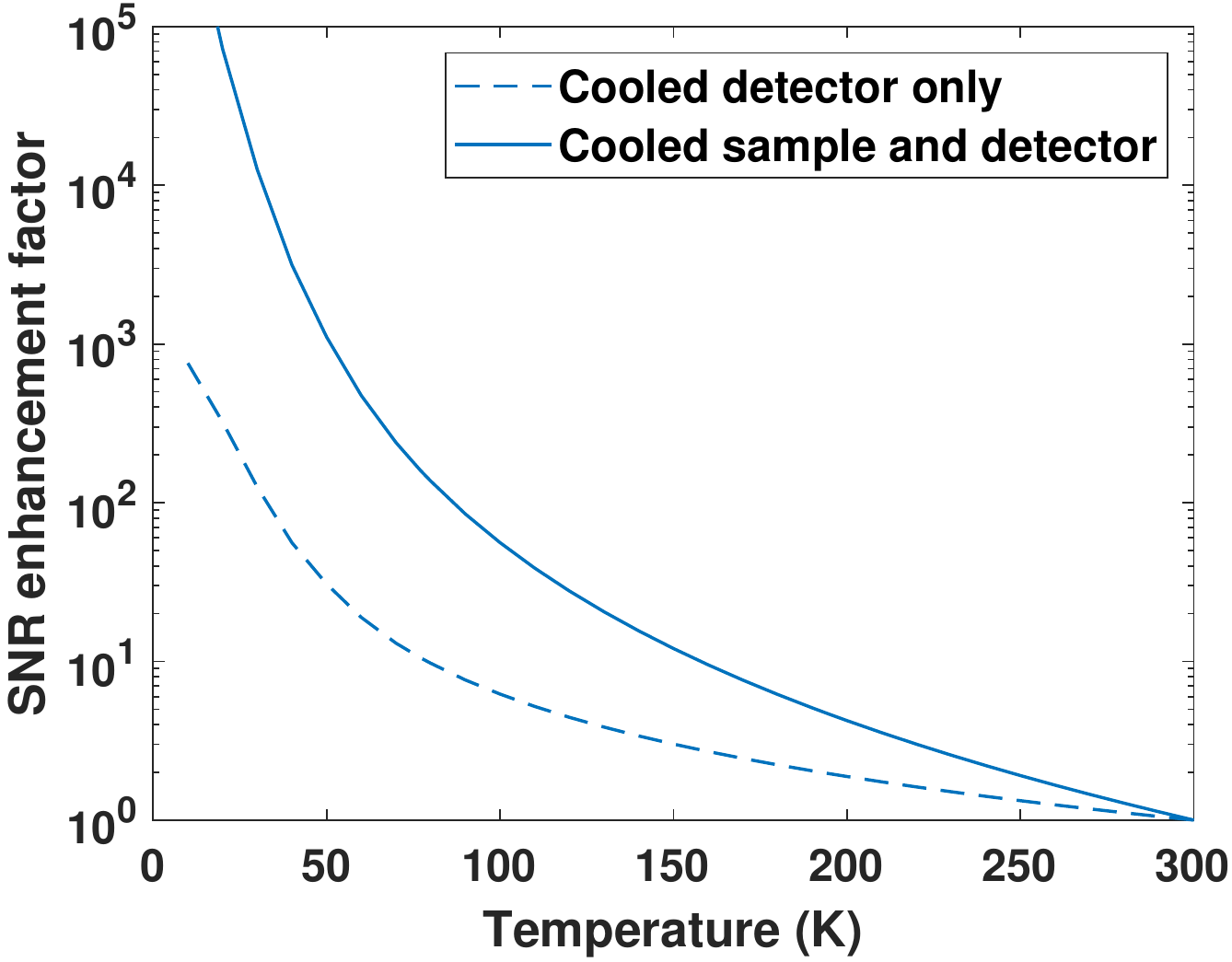}
    \caption{Estimated SNR enhancement factor as a function of experimental temperature.}
    \label{fig:snr_temp}
\end{figure}

We define the SNR enhancement factor as the ratio of $SNR_{e}$ at a specified temperature to its value at room temperature (300~K). Fig.~\ref{fig:snr_temp} summarizes the expected enhancement factor versus temperature for two cases: i) only the detector is cooled, and ii) both the detector and sample are cooled. For simplicity, we have assumed that receiver NF remains low over the entire temperature range; this is generally true for tuned probes using high-Q coils. The plot shows that even modest amounts of cooling can significantly increase $SNR_{e}$, and thus the number of scans required to obtain a given SNR.

Finally, we note that the relationship between $SNR_{e}$ and $SNR_{e,av}$, as defined by eqn.~(\ref{eq:snr_av}), also varies with sample temperature. This is because the relaxation times $T_{1}$ and $T_{2,eff}$ (which enter into the parameters $\alpha$, $\beta$, and $\delta$) are strongly temperature-dependent. In general, both decrease with temperature $\propto e^{E_a/(RT_s)}$ where $E_a$ is an activation energy parameter and $R$ is the molar gas constant~\cite{smith201514n}. Thus, obtaining the full benefits of sample cooling requires the SLSE pulse sequence parameters $t_w$ and $t_{SLSE}$ to be varied with temperature such that $\alpha$ and $\beta$ remain close to their optimal values.

\section{Tunable Probe Design}
\label{sec:probe_design}

\subsection{Series/Parallel-Tuned Probe} \label{subsec:series_parallel_tuned_probe}
One of the significant challenges of probe design for $^{14}$N NQR spectroscopy is the fact that the resonant frequencies are unique to each molecule and can vary over a broad range (1-5~MHz). As a result, tuned or matched probes must be capable of rapid and reliable re-tuning, ideally under software control. Broadband un-tuned probes provide an appealing alternative, but result in significant performance compromises in both transmit mode (reduced peak RF power handling capability) and receive mode (increased NF)~\cite{mandal2014ultra}. Thus, here we focus on digitally-tuned narrowband probes. In earlier work, Chen and Ariando have demonstrated such probes for NQR and low-field NMR that use networks of relays to program the capacitor values of a traditional two-capacitor impedance-matching network~\cite{nqr:cheng2017_vitamins,nqr:ariando2019autonomous}. 
However, two independent banks of relays are required to tune the probe, which greatly increases the complexity of the tuning algorithm. Impedance-matched probes at such low frequencies also suffer from limited bandwidth in transmit mode (which results in long pulse rise/fall times) and low voltage gain in receive mode (which makes the receiver NF sensitive to probe tuning errors). 

The usual solution, which is often found in low-field NMR systems such as well-logging tools~\cite{kleinberg1992novel}, is to use a single tuning capacitor $C_1$ in parallel with the sample coil $L_c$. This tuned (but not impedance-matched) design allows low-impedance switching power amplifiers to be over-coupled to the probe in transmit mode (which increases the transmit bandwidth) and also provides high voltage gain in receive mode (which reduces the effects of probe mistuning by providing low NF over a broad bandwidth, as discussed in the previous section). As a result, the AFE does not itself have to be cooled to maintain low NF over a wide range of probe temperatures, which greatly simplifies the receiver design. Moreover, a single bank of relays can be used to program $C_1$, resulting in straightforward tuning. However, the absence of impedance matching eliminates voltage gain during transmit, so high-voltage (HV) transmitters are required to obtain adequate $B_1$ amplitudes.

HV operation is problematic for electrical safety, and is particularly undesirable for portable devices. In addition, the figure of merit (FoM) for switching devices within the transmitter (e.g., MOSFETs or IGBTs) is a strongly-decreasing function of their voltage rating. Thus, it is beneficial to replace the high-impedance parallel-tuned load with a low-impedance series-tuned load that allows the use of lower transmit voltages for the same coil current (and thus $B_1$). However, series-tuning provides no voltage gain in receive mode, making it unsuitable for use with high-input-impedance pre-amplifiers. These issues can be addressed by using separately-tuned transmit and receive coils that are orthogonal to each other to minimize mutual coupling. However, the associated geometric constraints generally result in reduced coil sensitivity and fill factor, and may also complicate access to the sample during experiments. 

\begin{figure}
\centering
\includegraphics[width=0.95\columnwidth]{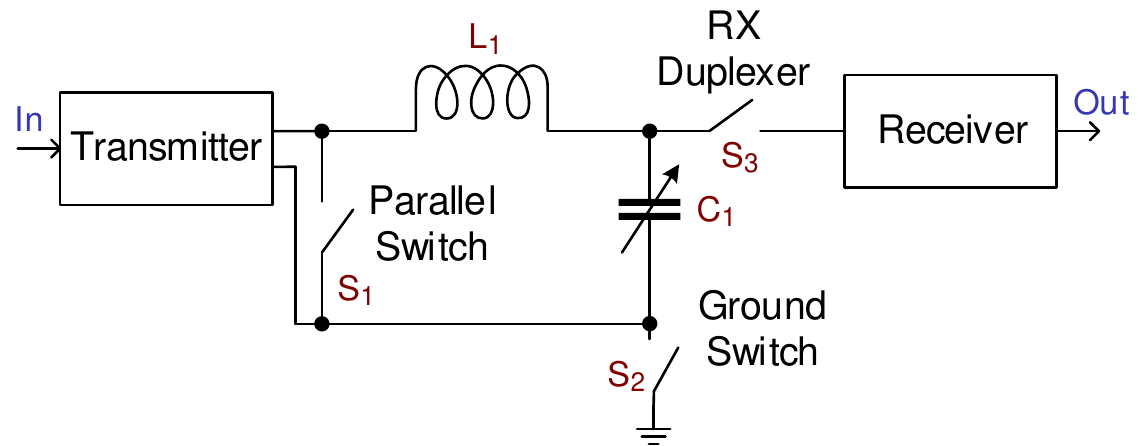}
\caption{Schematic of the dynamically reconfigurable tuned transmit-receive probe.}
\label{fig:coil_flipper_block_model}
\end{figure}

Fig.~\ref{fig:coil_flipper_block_model} presents an alternative approach in which high-performance gallium nitride field-effect transistors (GaN FETs) are used to rapidly switch a single-coil probe between two configurations during pulse sequences, namely series-tuned (for transmit) and parallel-tuned (for receive).

The reconfigurable tuning network uses three active bidirectional switches $S_1$, $S_2$, and $S_3$: all switches are ``off'' in transmit mode, and ``on'' in receive mode. GaN FET based switches are used due to their high switching FoM. For example, $S_2$ is realized using two back-to-back EPC2037 GaN FETs (EPC, El Segundo, CA), resulting in a typical on-resistance of $R_{on}=800$~m$\Omega$ and an off-capacitance of $C_{off}=3.25$~pF. The low value of $C_{off}$ minimizes its effect on resonant frequency during transmit mode. In receive mode, the relatively high impedance of the parallel resonant circuit minimizes the noise contribution of $S_2$ and $S_3$, but not $S_1$. Thus, $S_1$ uses devices with very low $R_{on}$. The same low-capacitance device as $S_2$ is suitable for $S_3$. Gate drive voltages for each switch are supplied by floating H-bridges with optically-isolated control inputs.

In the realized receiver, a fourth switch (not shown) with series resistor is routed in parallel with $S_1$; the resulting current shunt decreases the ringdown time at the end of transmit pulses. The effect is realized by selecting a resistor which critically damps oscillations of the resonant network at the end of transmit pulses. The optimal value of this resistor is strongly dependent on coil $Q$, and thus both $\omega_0$ and $T_c$. To simplify the system, we used a fixed resistor optimized for nominal values of these parameters ($2\pi\times 2.5$~MHz and 300~K, respectively).

\subsection{Optimization of the Sample Coil Geometry}
\label{subsubec:coil_flipper_spice}
The available probe tuning range is heavily dependent on the parameters of the physical spectrometer. Selection of receiver coil dimensions, its position relative to the sidewalls of the enclosure, and distance to the receiver are the dominant influences. In addition, secondary effects include parasitic and stray effects in the realized PCB layout, component selection, and conductive enclosures. The potentially detrimental effects of nearby shielding (or lack thereof) coupled to the highly sensitive network are analyzed later.

\begin{figure}
\centering
\includegraphics[width=0.90\columnwidth]{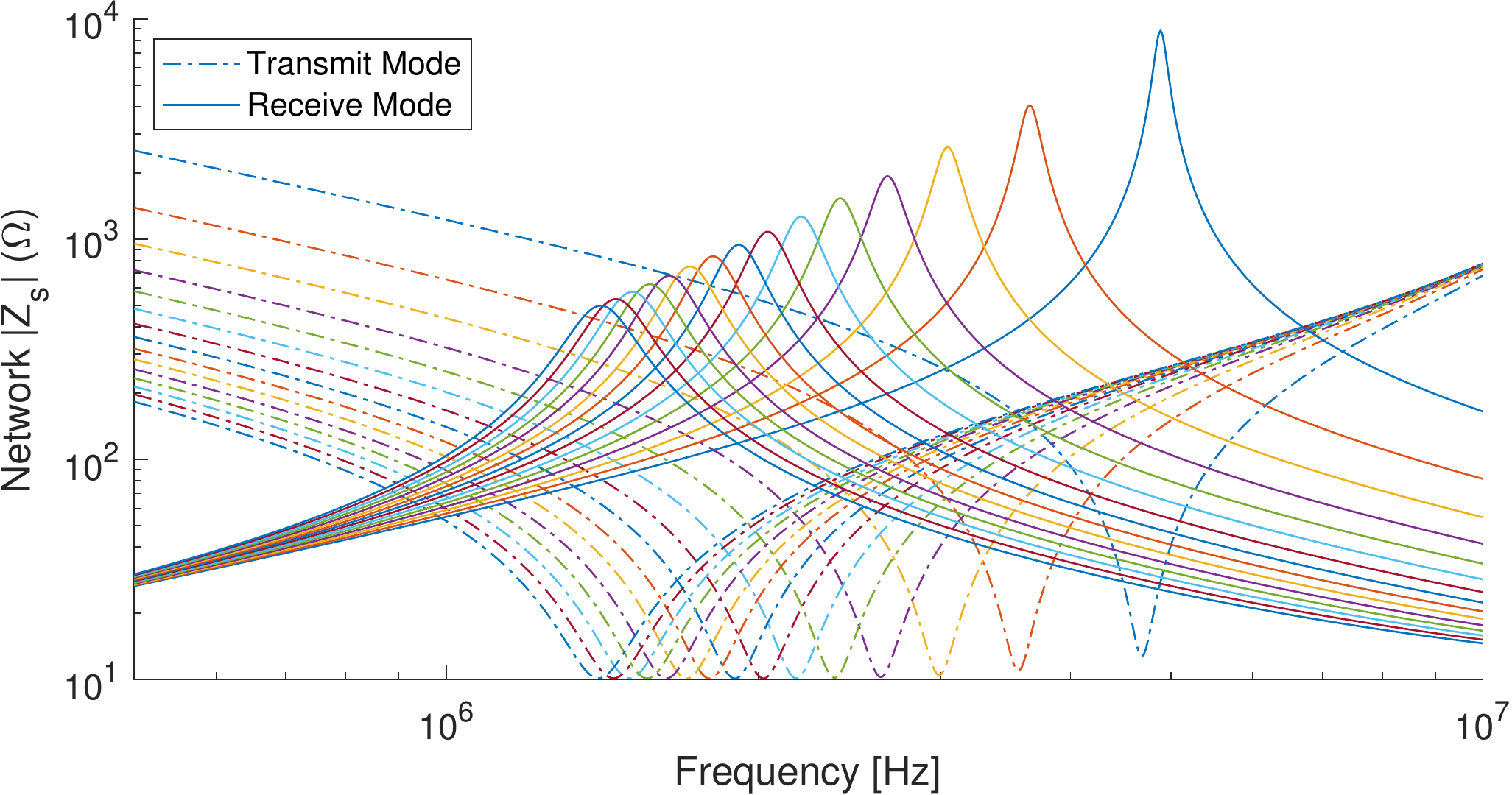}
\caption{Impedance sweep shown for transmit (-$\cdot$-) and receive (---) modes with a typical coil and stepped values for $C_{p}$.}
\label{fig:coil_flipper_spice_sim_tuning_txrx}
\end{figure}

Fig.~\ref{fig:coil_flipper_spice_sim_tuning_txrx} presents simulated impedance $\left|Z_{\rho}\right|$ of the reconfigurable resonant network in both transmit and receive modes for a typical sample coil. Each of fifteen positions displayed in the figure are obtained by varying the tuning capacitance $C_{t}\in [100~{pF}, 1500~{pF}]$ in increments of $100$~pF, while $L_{c}$ is fixed at 8.36~$\mu$H with $R_{s}=$370~m$\Omega$ and $C_{p}=10$~pF. 

\begin{figure}
\centering
\includegraphics[width=0.925\columnwidth]{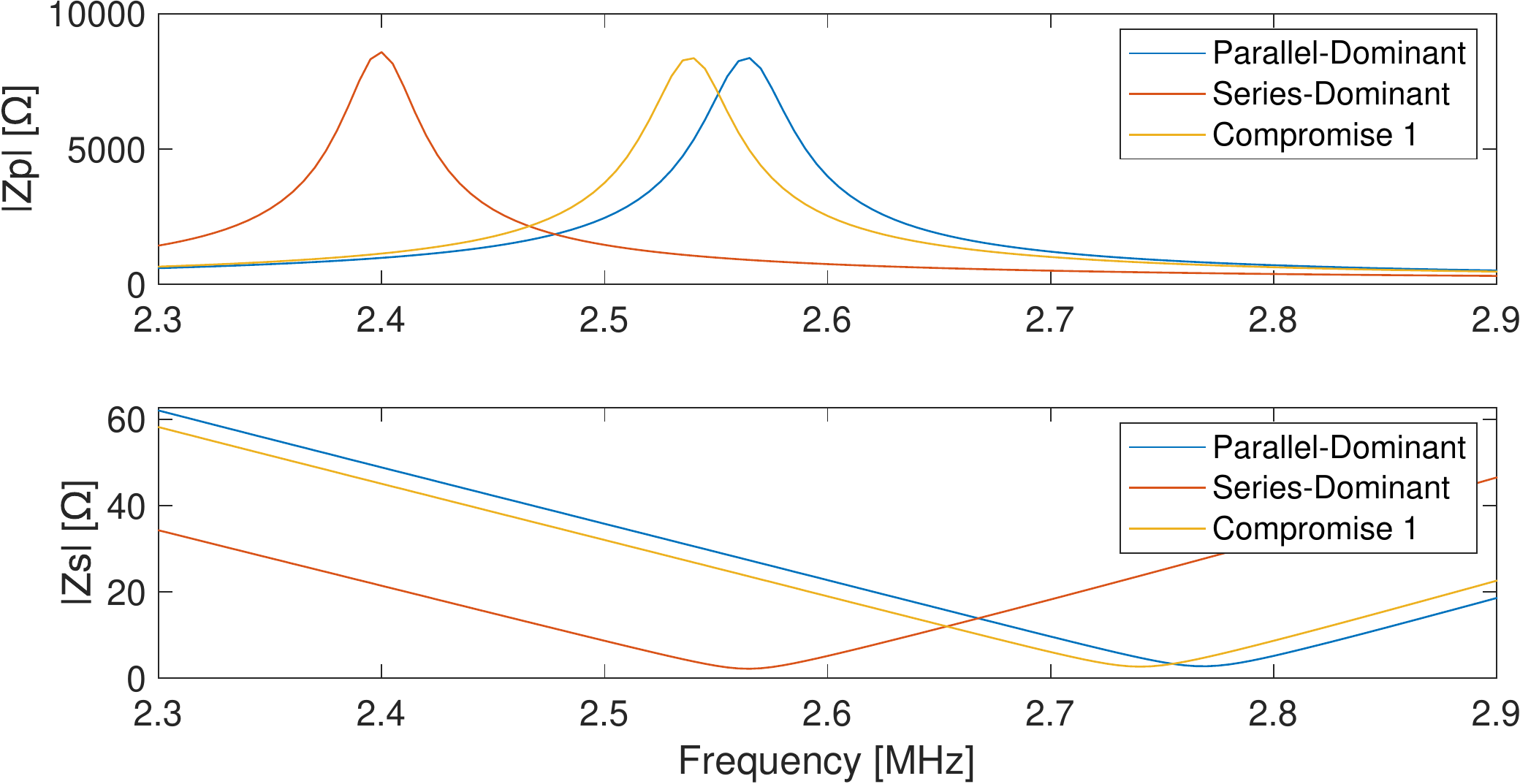}
\caption{Measurement of network impedance in parallel and series modes while adjusting tuning capacitance.}
\label{fig:coil_flipper_mistune_real_data}
\end{figure}

Fig.~\ref{fig:coil_flipper_spice_sim_tuning_txrx} figure predicts a good match (within 40~kHz) between the resonant series- and parallel-tuned configurations over the entire tuning range. However, this model is only accurate when the coil is in free space away from nearby conductors. In a shielded enclosure (necessary to minimize RFI), capacitive coupling to the walls of the box amplifies the mismatch between resonant frequencies and thus makes it difficult to optimally tune both modes simultaneously. The effect is visualized in Fig.~\ref{fig:coil_flipper_mistune_real_data}. Using the same coil as for simulation and tuning to 2.564~MHz~\footnote{One of the $^{14}$N resonant frequencies in acetaminophen}, tuning for optimum receive performance (a ``parallel-dominant'' approach) results in a somewhat high corresponding series mode impedance 27~$\Omega$. Tuning for optimum transmit performance (a ``series-dominant'' approach) nearly eliminates the beneficial passive gain of the tuned network for the receiver, though the series impedance drops to 2.2~$\Omega$. Moving the coil to a safe distance away from nearby shielding and enclosing surfaces reduces the effective tuning gap and enables a compromise position which yields $>70$\% of peak receive mode $\left|Z_{\rho}\right|$ and agrees with simulations. 

Detector coil designs must balance between maximizing the sensitivity function (which is proportional to $V$ and $N/l$) and minimizing $C_{p}$ (which determines the gap between series- and parallel-mode resonant frequencies). A large sample volume is desirable for greater nuclear spin population but requires relatively large (thus long) coils for which $C_{p}$ becomes large. Sensitivity to $C_{p}$ is significantly alleviated by reducing the inductance of the detector coil via either the turns density or the total number of turns. 

Various coil designs for the cryogenic detector were wound by hand and experimentally characterized using a vector network analyzer (VNA) to evaluate the effects described above. The VNA converts two-port measurements into the equivalent series R-L model $Z = R_s + sL_s$, which would be accurate if the device under test was truly just an inductor with series resistance. However, this does not account for parasitic capacitance. Approximating the latter as a parallel capacitance $C_p$, consider the values reported by the network analyzer to be $Z = R_{net} + sX_{net}$. The real values of coil resistance $R_s$ and inductance $L_s$ are then related by
\begin{equation}
R_{net} + sX_{net} = \frac{(sL_s + R_s)s^{-1}C_p^{-1}}{sL_s+R_s+s^{-1}C_p^{-1}}.
\end{equation}
The system was reduced to two variables by approximating inductance using a low-frequency measurement. An estimator for $C_p$ was then found as that of the series R-L-C circuit with self-resonant frequency $f_{sr}$, such that $C_p = 1/\left(4\pi^2 f_{sr}^2 L_s\right)$.
The model was used to compute actual coil parameters by iterative error minimization. Table~\ref{tab:rx_coil_sampling} summarizes the estimated R-L-C parameters for some of the test coils used in the setup.

\begin{table}
\begin{center}
\caption{\label{tab:rx_coil_sampling}Measurement parameters of four solenoidal test coils for the cryo-system. Here TPC denotes the turn density, i.e., turns/cm.}
\begin{tabular}{@{} c c c c c c @{}} \toprule
\multirowcell{2}{\textbf{Coil}\\ \textbf{Name}} & \multicolumn{5}{c}{\textbf{Parameters}} \\ \cmidrule{2-6}
                 &  \textbf{N\sub{turns}}  & \textbf{TPC} & \textbf{R\sub{@1~MHz} [$\mathbf{\Omega}$]}  &   \textbf{L\sub{s} [$\mu$H]}   & \textbf{C\sub{p} [pF]}   \\ \midrule
        A        &  46  &     6       &  0.823  &  22.5  &  9.21 \\ \midrule
        C        &  32  &     5.5       &  0.494  &  13.1  &  8.72 \\ \midrule
        D        &  32  &     5       &  0.534  &  11.6  &  8.67 \\ \midrule
        E        &  27  &     4       &  0.369  &   8.3  &  9.22 \\ \midrule
\end{tabular}
\end{center}
\end{table}

Table~\ref{tab:rx_coil_sampling} suggests that coupling to nearby metallic objects is the dominant effect on the measured $C_{p}$, with stray capacitance between the turns contributing less than expected. Thus, increasing the turns packing ratio $N/l$ is a convenient option for increasing coil sensitivity since it has limited effect on $C_{p}$. However, for a fixed length (fixed $V$) it also increases $R_{coil}$ (thus source noise) and $L_{s}$. Cryogenic cooling reduces the impact of $R_{coil}$ on measurement sensitivity if a high turns ratio is desired; that said, it is beneficial to have relatively small coils with low inductance. Minimum inductance coils reduce the impact of small variances $\Delta C$ on the resonant frequency of the tuned detector, thus improving robustness against parasitic effects and allowing a greater degree of control over tuning.

The test results summarized in Table~\ref{tab:rx_coil_sampling} were used to select the optimum coil geometry for our setup (design ``A'' in this case). Models for automated optimization of the detector coil could be developed using a cost function based on measurement sensitivity, but were not pursued during this study.

\subsection{Optimization of the Tuning Network}
\label{subsubec:coil_tuning}

\begin{figure}
\centering
\includegraphics[width=0.6\columnwidth]{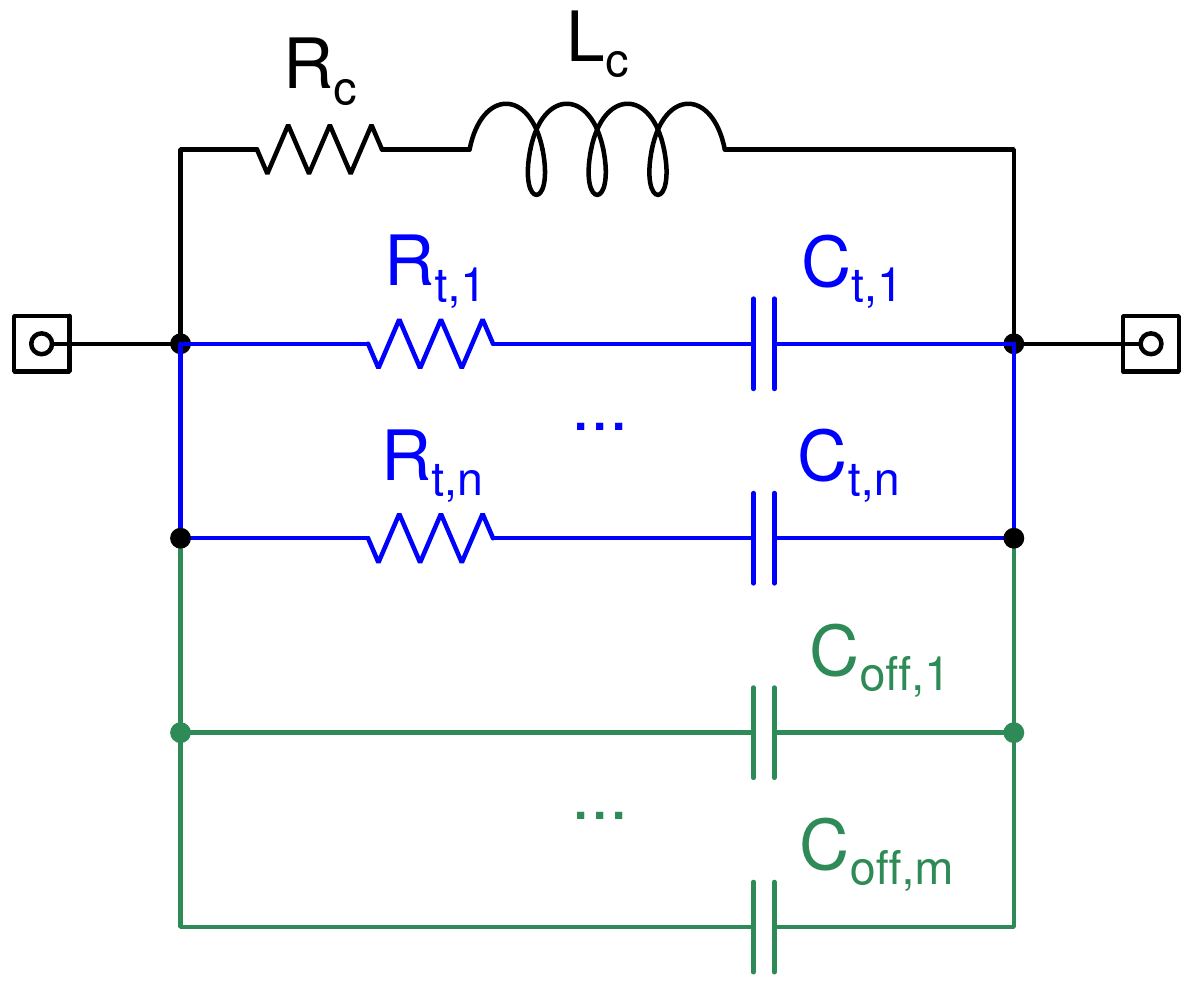}
\caption{The proposed switch-based tuning network. There are a total of $N$ tuning capacitors with series switches, of which $n$ are in the `on' state (blue) and $m=N-n$ are in the `off' state (green).}
\label{fig:rlynet_simp1}
\end{figure}

Options for realizing an electronically-controllable tuning capacitor include i) voltage-controlled capacitors (varactors), and ii) arrays of fixed capacitors $C_{t,i}$, $i\in [1,N]$, each enabled/disabled by a switch. Varactors enable continuous tuning and can be cryogenically cooled along with the coil, but suffer from limited breakdown voltages (which results in a severe trade-off between power handling capability and tuning range). Thus, we opt for a switch-based tuning network (Fig.~\ref{fig:rlynet_simp1}). Our choice of switching device is guided by their FoM, which is defined as $R_{on}\times C_{off}$ where $R_{on}$ is the resistance in the `on' state and $C_{off}$ is the capacitance in the 'off' state. Since $R_{on}$ degrades SNR by contributing noise while $C_{off}$ limits the tuning resolution, it is important to pick devices with minimal FoM. As in our earlier work~\cite{nqr:cheng2017_vitamins}, we used miniature electromechanical switches (reed relays) since they offer both low FoM ($\sim 100\times$ lower than semiconductor switches) and relatively fast switching times (100~$\mu$s -- 1~ms).

The number of tuning capacitors, $N$, was chosen to ensure complete frequency coverage over a range $1-5$~MHz, which is sufficient for most $^{14}$N sites in pharmaceutical compounds. By complete coverage, we imply that the spacing $\Delta f_0$ between available tuning frequencies is always smaller than the bandwidth $f_{0}/Q$ of the tuned network, i.e., $\Delta f_0 < f_{0}/Q$, such that arbitrary RF frequencies can be chosen for NQR experiments. To determine the minimum value of $N$, we developed a model of the tuning network with an arbitrary number of capacitors $n<N$ in the on-state. The latter are treated as ideal capacitors $C_{t,i}$ in series with small resistances $R_{t}=R_{on} + ESR_{C}$ where $ESR$ is the equivalent series resistance of each tuning capacitor. In addition, the $m=N-n$ off-state capacitors are approximated as $C_{off}$, the open-state capacitance between the relay contacts. The quality factor of the network, which determines the probe bandwidth, is then estimated as $Q\equiv X(\omega)/R(\omega)$ where $X(\omega)$ and $R(\omega)$ are the imaginary and real components of the probe impedance $Z(\omega)$, respectively.

\begin{figure}
\centering
\includegraphics[width=0.8\columnwidth]{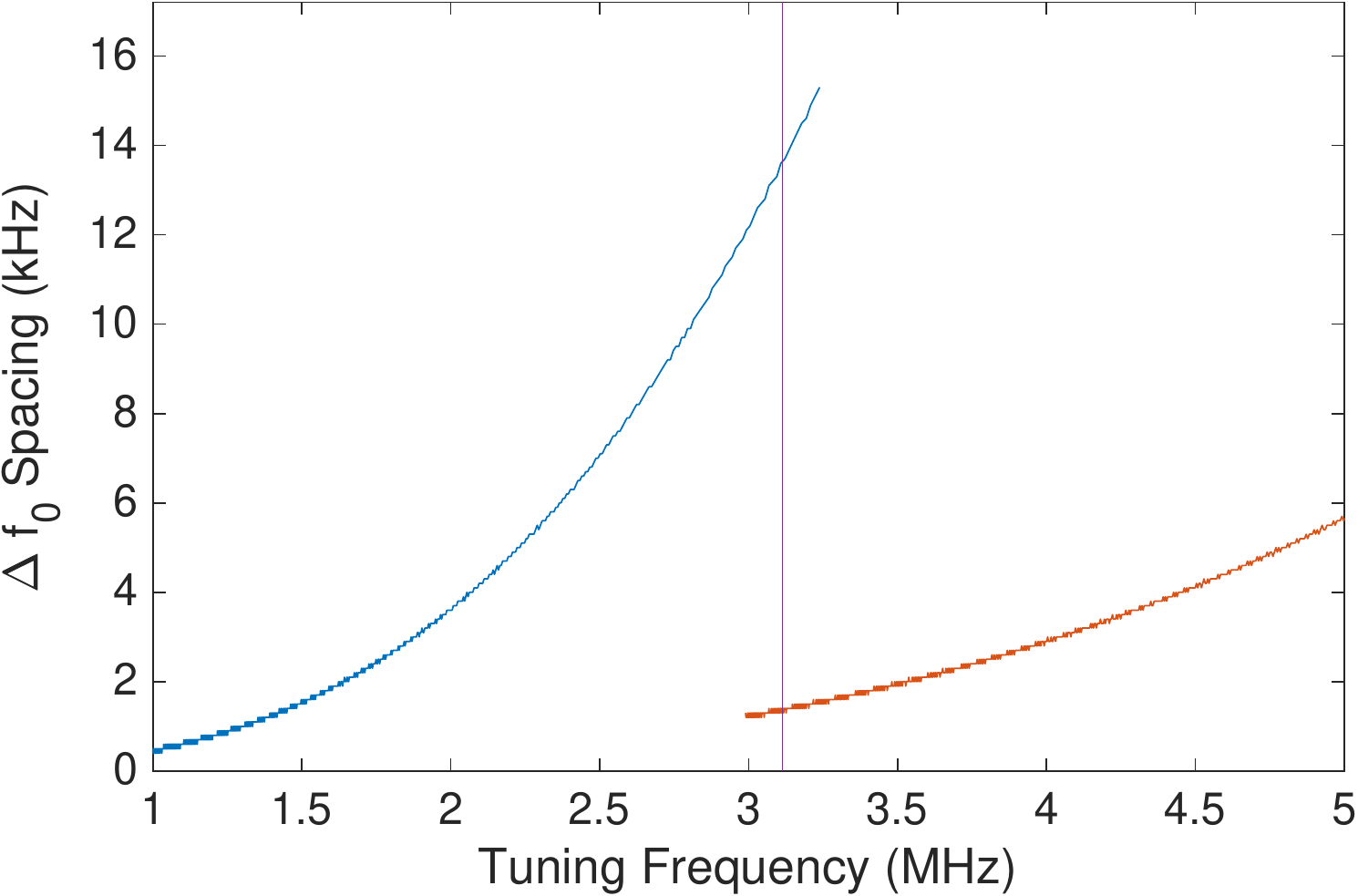}
\caption{Simulated spacing $\Delta f_0$ between tuning frequencies for a 20-channel dual-bank tuning network with an additional fixed capacitance (100~pF) in parallel with the switched capacitors. The following relay parameters were assumed: $R_{on}=0.1$~$\Omega$, $C_{off}=0.1$~pF.}
\label{fig:relay_freqdistance_20ch}
\end{figure}

Since $Q$ increases with frequency, smaller step sizes are required as $f_{0}$ increases. Thus, using a single binary-weighted array of tuning capacitors, as in previous work~\cite{nqr:cheng2017_vitamins}, is inefficient since it results in a fixed value of $\Delta f_0$ that must be set based on the worst case (i.e., to ensure coverage at the highest tuning frequencies). Instead, we use two separate banks of binary-weighted capacitors with different values of the least significant bit (LSB). Specifically, we use two sets of $N_1=10$ capacitors, one with values in the range $C_{t,i} \in 5~\mathrm{pF}\times \lbrace 1,2,\,...\,,2^{10}-1 \rbrace$ and the other in the range $C_{t,i} \in 1~\mathrm{pF}\times \lbrace 1,2,\,...\,,2^{10}-1 \rbrace$. The resulting frequency spacing for the chosen sample coil is shown in Fig.~\ref{fig:relay_freqdistance_20ch}; the worst case value of $\sim$14~kHz is adequate at room temperature but too large for cryogenic operation due to the higher value of $Q$.

\begin{figure}
\centering
\includegraphics[width=0.9\columnwidth]{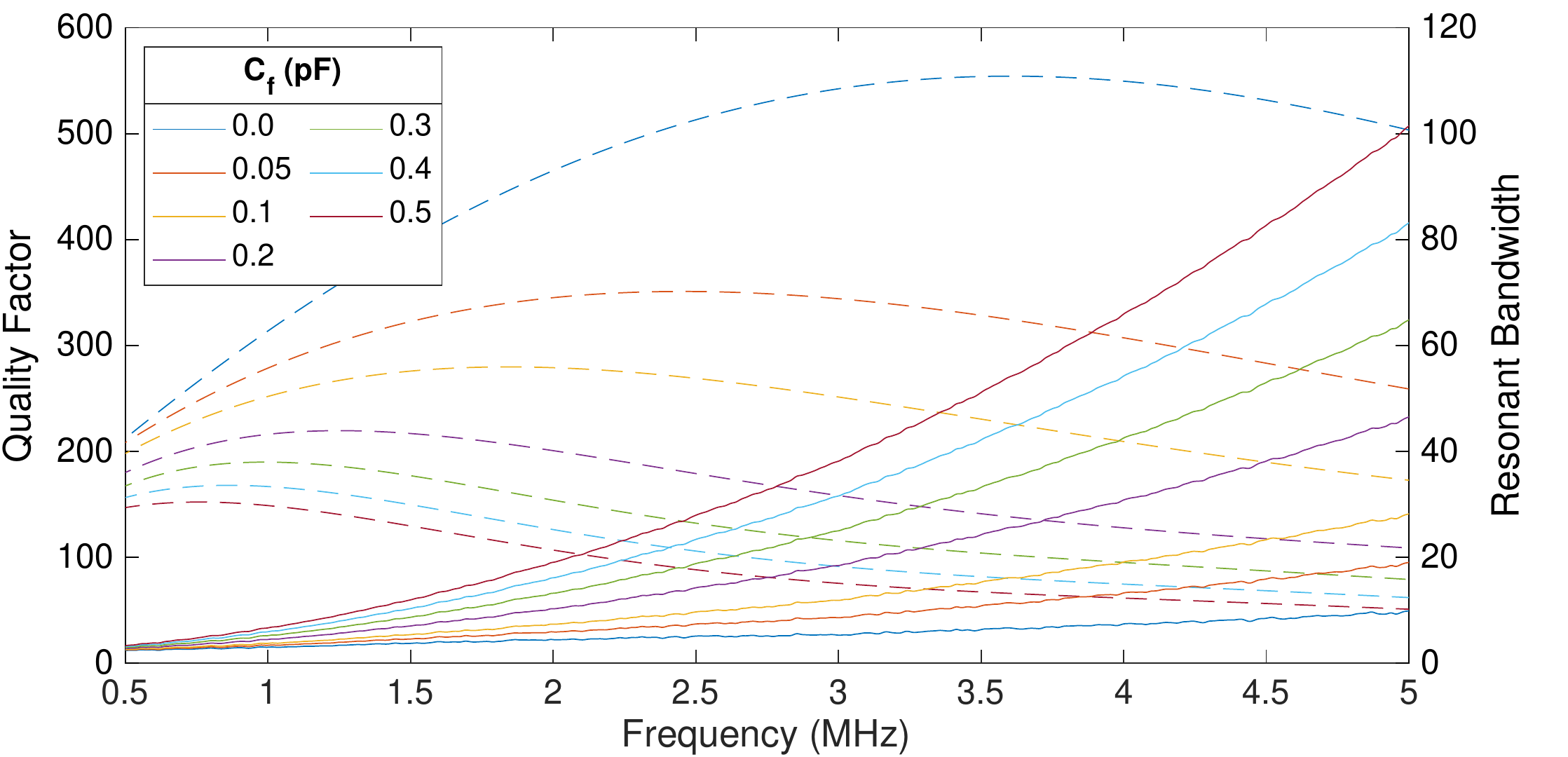}
\caption{Simulated quality factor ($Q_{eff}$) and bandwidth of the damped network at 77~K for various values of the feedback resistor $C_{fb}$.}
\label{fig:Q_damping_bandwidth_LN2T}
\end{figure}

To ensure adequate coverage at cryogenic temperatures, one option is to simply increase the resolution of the tuning network to reduce $\Delta f_0$, i.e., by reducing the LSB capacitance and adding more tuning channels. However, the off-state relay capacitance $C_{off}=0.1$~pF places a lower bound on the achievable step size; the tuning function becomes non-monotonic when the LSB capacitance becomes comparable to $C_{off}$. Instead, we relax the requirement on $\Delta f_0$ by using a noiseless feedback network to increase the probe bandwidth, i.e., reduce its effective $Q$ without adding noise~\cite{mandal2014ultra}. The method uses a capacitor of value $C_{fb}$ in series with a $90^{\circ}$ phase shifter, for instance an integrator with transfer function $H(j\omega) = A/(j\omega \tau)$. The phase shift effectively generates a resistance by shifting the phase of the capacitor's impedance into the real plane, but does not add noise since $C_{fb}$ is (ideally) noiseless. The capacitor thus behaves like a grounded resistor of value
\begin{equation}
R_{damp} = \frac{\tau}{A C_{fb}}
\end{equation} 
in parallel with the tuning network. The effective $Q$ of the network is then given by
\begin{equation}
Q_{eff} = \frac{Q}{\left(\omega_0L_cQ/R_{damp}\right)+1},
\end{equation}
where $Q\gg 1$ is the original (un-damped) quality factor of the tuning network. Fig.~\ref{fig:Q_damping_bandwidth_LN2T} summarizes the quality factor ($Q_{eff}$) and bandwidth ($f_0/Q_{eff}$) of the damped network for various values of $C_{fb}$ and typical integrator parameters ($A$,$\tau$). The figure shows that a relatively small capacitance ($C_{fb}\approx 0.1$~fF) suffices to provide sufficient bandwidth over the entire tuning range. Larger values of $C_{fb}$ are not recommended since the voltage gain $Q_{eff}$ of the damped probe also decreases, eventually resulting in higher receiver NF as predicted by eqn.~(\ref{eq:rx_noise}). 

\section{A Custom Broadband NQR Receiver} 
\label{sec:nqr_receiver_2} 

\begin{figure}[htbp]
\centering
\includegraphics[width=0.85\columnwidth]{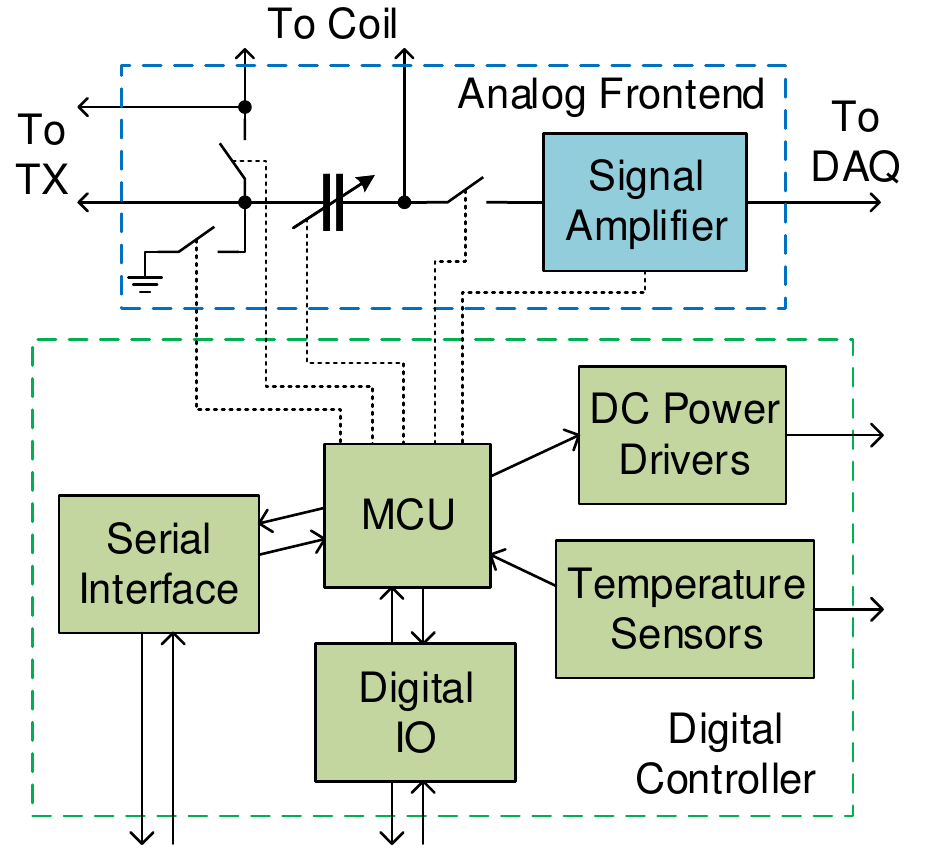}
\caption{Block diagram of the proposed NQR receiver, which consists of an analog front-end and a digital controller.}
\label{fig:rx2_0_receiver_overview}
\end{figure}

The receiver is organized into two primary components: an analog front-end (AFE) and a digital controller. The AFE consists of the signal path between the detector coil and measurement data acquisition; the latter utilizes a commercial benchtop NMR spectrometer (Magritek Kea2). The digital controller facilitates experiment sequencing and inter-experiment preparation, including tuning of the resonant detector, monitoring ambient climate, control over polarization enhancement, and enabling/disabling the analog receiver. Digital input/output (I/O) pins are also used to check status of battery charge and pause the experiment if any battery packs are low. These components are shown in the block diagram of Fig.~\ref{fig:rx2_0_receiver_overview}.

\subsection{Analog Front-end Design and Circuit Performance}
The AFE consists of two main components: the duplexer and the preamplifier. The duplexer uses FET switches to enable broadband operation. It was initially designed using cascode topology GaN FETs (Transphorm, USA) to tolerate a greater (650~V) input amplitude, at the cost of increased input capacitance in the off state. However, the additional capacitance resulted in significant de-tuning, even for small coils. Thus, the circuit was redesigned using smaller 100~V rated GaN FETs (EPC, USA) to minimize off-resonance effects in the tuned detector when it is mated to the receiver. 


Apart from the tuned detector, the AFE has a broadband frequency response. The input stage of the preamplifier uses a silicon JFET configured as a source follower to ensure both low voltage and current noise and high input impedance~\cite{mandal2014ultra}. A small 1:4 step-up transformer is used to realize nearly noiseless broadband voltage gain before the input JFET, thus further improving the noise figure (NF). The simulated worst-case NF (assuming noiseless feedback damping of the probe) is $<0.01$~dB over the entire temperature and frequency ranges of interest; such excellent performance is due to the high passive voltage gain $Q_{eff}\gg 1$ provided by the tuned probe.

As noted in prior art~\cite{glickstein-automated-PENQR}, such broadband preamplifier designs are susceptible to environmental noise and transformer self-resonance. Fortunately, the feedback damping network (which is mainly used to increase probe bandwidth, as described in the previous section) also mitigates these effects. The $90^{\circ}$ phase shifter required by the damping circuit was realized using an op-amp integrator. The gain of various stages in the receiver was adjusted to compensate for the reduced gain resulting from the damping circuit. The gain and width of the passband are only mildly impacted by these changes.

\subsection{Mixed Domain Isolation}


Several means of mixed signal domain isolation were implemented in the spectrometer. The intent is to achieve good mixed signal noise immunity by reducing the coupling between domains when passing control signals. Note that these techniques are not effective in isolation. The digital controller and analog front-end were powered by separate DC supplies, with a star ground topology used to avoid coupling of digital switching noise into the amplifier. The receiver uses latching relays to tune the detector while maintaining electrical isolation from the digital controller. The DC supply (battery pack) used by the relay coils is enabled through a momentary digital switch which is disconnected while not in use. 

Four floating gate drivers were used to implement the detector described in Section~\ref{subsec:series_parallel_tuned_probe}. Each driver is supplied by an isolated battery pack and the state of each gate is controlled through optical isolators. Note the use of such isolators is not a complete solution, since they do not prevent digital switching noise from crossing into the analog domain of the preamplifier without careful filtering of the input signals which drive the LED in each isolator. Furthermore, the floating supplies create a new challenge in ensuring noise immunity. The lack of a defined ground in each duplexer supply and control circuit with respect to the spectrometer ground makes each duplexer highly susceptible to noise coupling from external sources. Thus, these supplies are well shielded inside the spectrometer, and wires between each duplexer battery pack and the receiver PCB are kept minimal in length. 

Communication between the digital controller and an external computer uses a serial port. A USB signal isolator using monolithic air core transformers is used to electrically isolate the computer from the spectrometer. This device is primarily intended to isolate the computer from damage resulting from unanticipated circuit failures. Noise at the communication frequency is not isolated, though this occurs at 480~MHz and is high pass filtered by the air core transformers to prevent coupling low frequency noise from the external computer system. 

\subsection{Fully-Assembled Receiver}

\begin{figure}
\centering
\includegraphics[width=0.85\columnwidth]{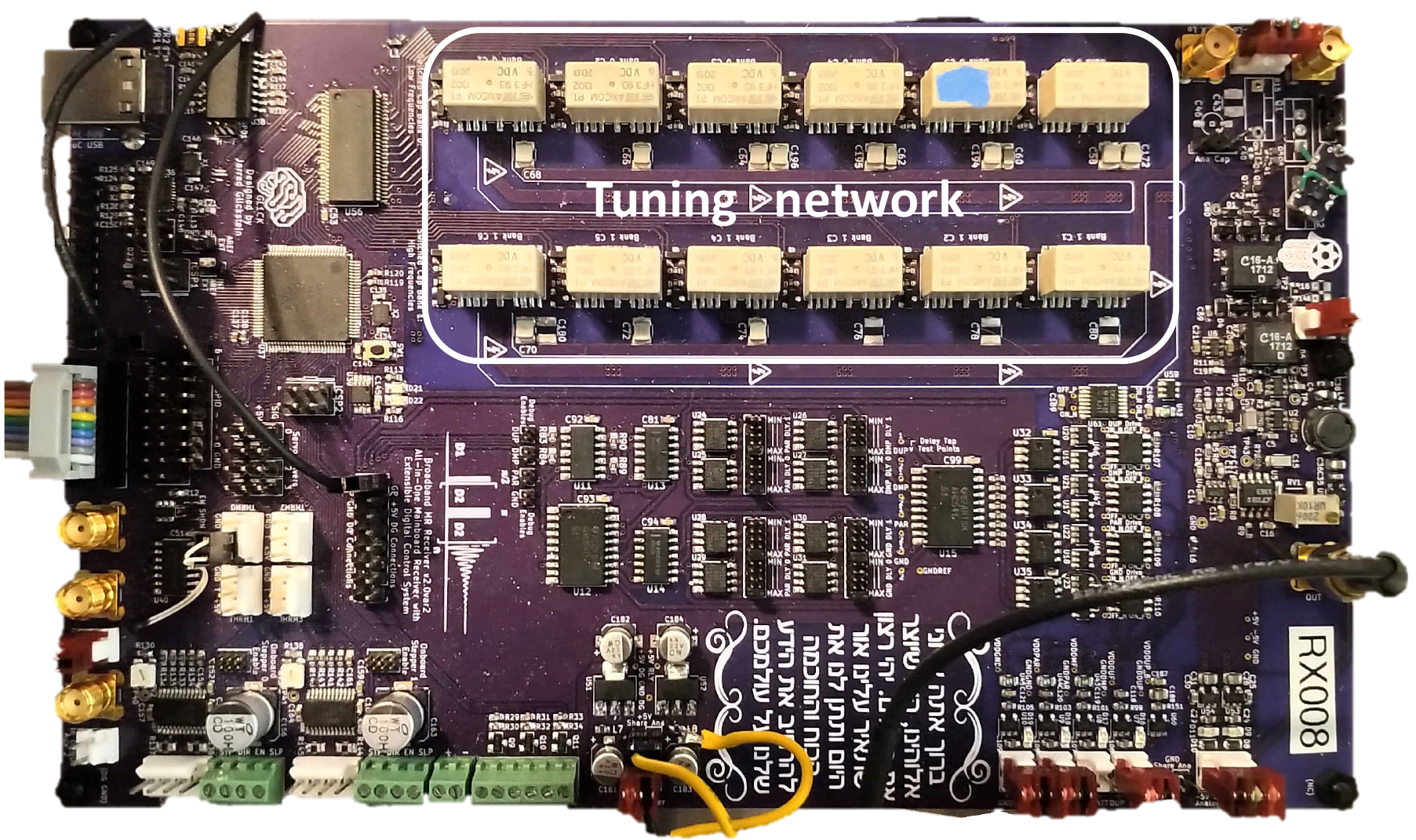}
\caption{Completely assembled receiver in the experimental setup at room temperature. The tuning network is labeled.}
\label{fig:nqr_receiver_v2_assembled}
\end{figure}

An assembled receiver board is shown in Fig.~\ref{fig:nqr_receiver_v2_assembled}. General purpose digital and analog IO near the bottom of the board allow versatile adaptation of the digital controller to evolving design intent. Flexible transmitter and coil connectivity (at top left in Fig.~\ref{fig:nqr_receiver_v2_assembled}) provide similar functionality. 

\section{A Broadband RF Pulse Transmitter} \label{sec:nqr_transmitter_v2}

RF pulses are delivered by a broadband class-D amplifier (a H-bridge) similar to that used in our previous work~\cite{jglickstein-ms}. However, the present design is series-tuned in transmit mode, which results in significantly better power transfer from the amplifier~\cite{sato2014design,otagaki2020development}. Specifically, the low impedance of the series-tuned coil ($\approx R_{c}$ on resonance) is now relatively well-matched to the low output resistance of the amplifier ($\approx 2R_{on}$, where $R_{on}$ is the on-state resistance of the switches in the H-bridge). As a result, the system no longer requires a HV power supply to deliver sufficient current to the coil.

For improved power efficiency, the H-bridge was implemented using discrete GaNFETs (Transphorm, USA). The board layout and components was designed to support pulse currents up to 5~A. However, our experiments used lower current levels. Specifically, a $\approx 9.5$~V supply was used to develop typical pulse amplitudes of $1$~A\sub{pk}. Note that the required supply voltage is unusually high for a tuned load due to off-resonance effects (see Fig.~\ref{fig:coil_flipper_mistune_real_data}); the coil was always tuned to be on-resonance in the parallel configuration to take full advantage of the high Q in receive mode.

\begin{figure}
\centering
\includegraphics[width=0.85\columnwidth]{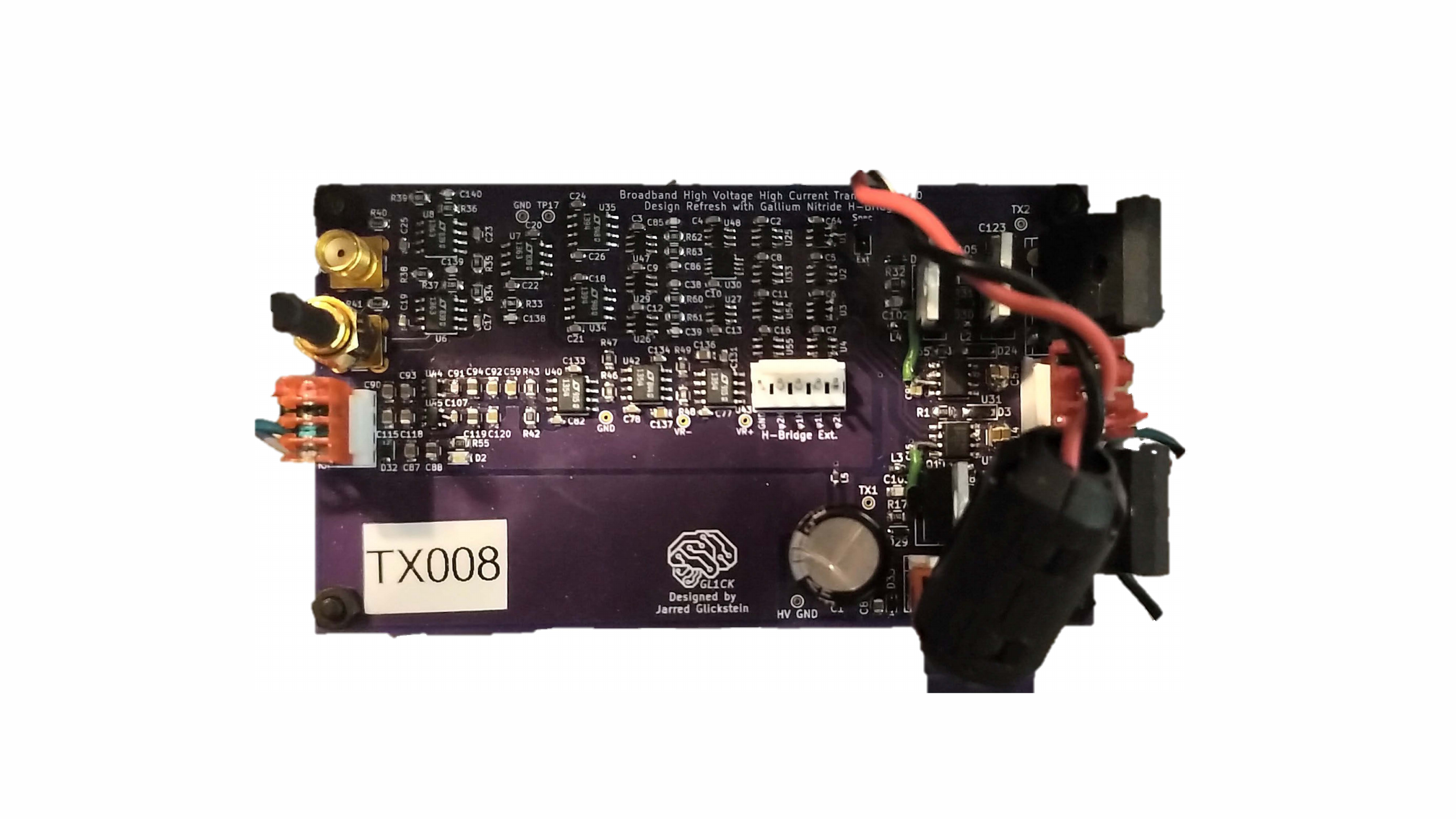}
\caption{Completely assembled transmitter in the experimental setup at room temperature.}
\label{fig:nqr_transmitter_v2_assembled}
\end{figure}


\section{System Implementation and Test Results}
\label{sec:nqr_gen2_spectrometer}

\subsection{Power Supplies}
\label{sec:power_supplies}

The spectrometer was powered using battery packs to eliminate power supply noise and ripple. Two variations were utilized, a 4-cell split rail supply with output voltage $\pm$7.4~V and a 2-cell single rail supply with output voltage 3.7~V. The battery packs used 18650 size lithium-ion cells because they offer reasonably large capacity and are readily available at reasonable prices due to their popularity. The nominal cell voltage is 3.7~V and the typical fully charged voltage is 4.2~V. Each cell has dimensions $18\times 65$~mm\ ($H \times D$). The battery packs implement two state charge monitors (``OK'' and ``RECHARGE'') which interface with the receiver digital controller using an optically isolated output.
Isolation is required not only to maintain noise immunity, but to maintain separation of floating gate drivers from the receiver ground. In the event of a low battery signal, the receiver indicates a fault condition using a colored LED, signals to a control program over its serial interface, and halts data acquisition. In addition, a sealed lead acid battery provides the nominally\ 12~V unregulated supply to switch the H-bridge in the transmitter.


\subsection{Design of the Cryogen Tank}
\label{sec:second_generation_cryotank}

The cryogen tank (i.e., cryotank) has interior dimensions of $30.5\times 71 \times 30.5$~cm ($H\times W \times D$) and was made of welded aluminum panels to ensure mechanical robustness. Specifically, it was constructed using 6.35~mm 6061 aluminum sheets. However, the high thermal conductivity of aluminum requires the addition of an outer insulating layer. Wood is inexpensive, easy to assemble, and a poor conductor of heat due to its high porosity. The outer layer of the cryotank was therefore produced from birch plywood, which has a low thermal conductivity of 0.11~W/mK~\cite{crc_handbook}. The structure was then insulated using closed cell insulation foam packed within a 8.9~cm gap between the aluminum sheets and outer wooden walls.


Despite the insulation, testing revealed that the outer walls were being cooled below the dew point of the local environment, causing water vapor to condense in the wood. Since water is an excellent conductor of heat (0.6~W/mK), further enhanced when frozen (1.6~W/mK), such condensation limits the operating lifetime of the tank. Specifically, the tank was measured to have a useful operating lifetime (before refilling) of 3-4~hours. While this is sufficient for most experiments, revised tank designs may nevertheless replace the wood with a plastic (e.g., polyvinyl chloride) to trade off somewhat higher thermal conductivity for considerably lower porosity.

\begin{figure}
\centering
\includegraphics[width=1\columnwidth]{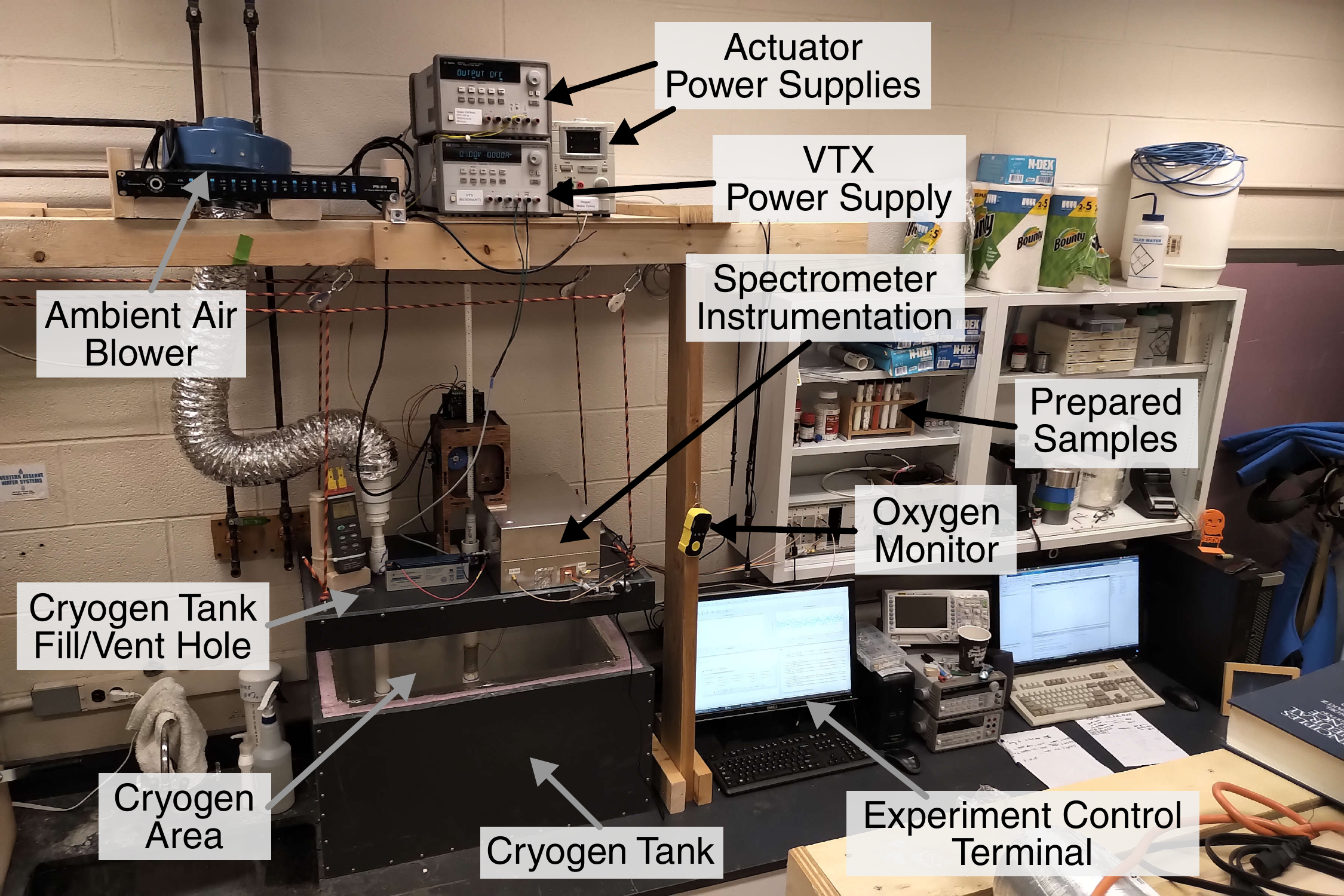}
\includegraphics[width=1\columnwidth]{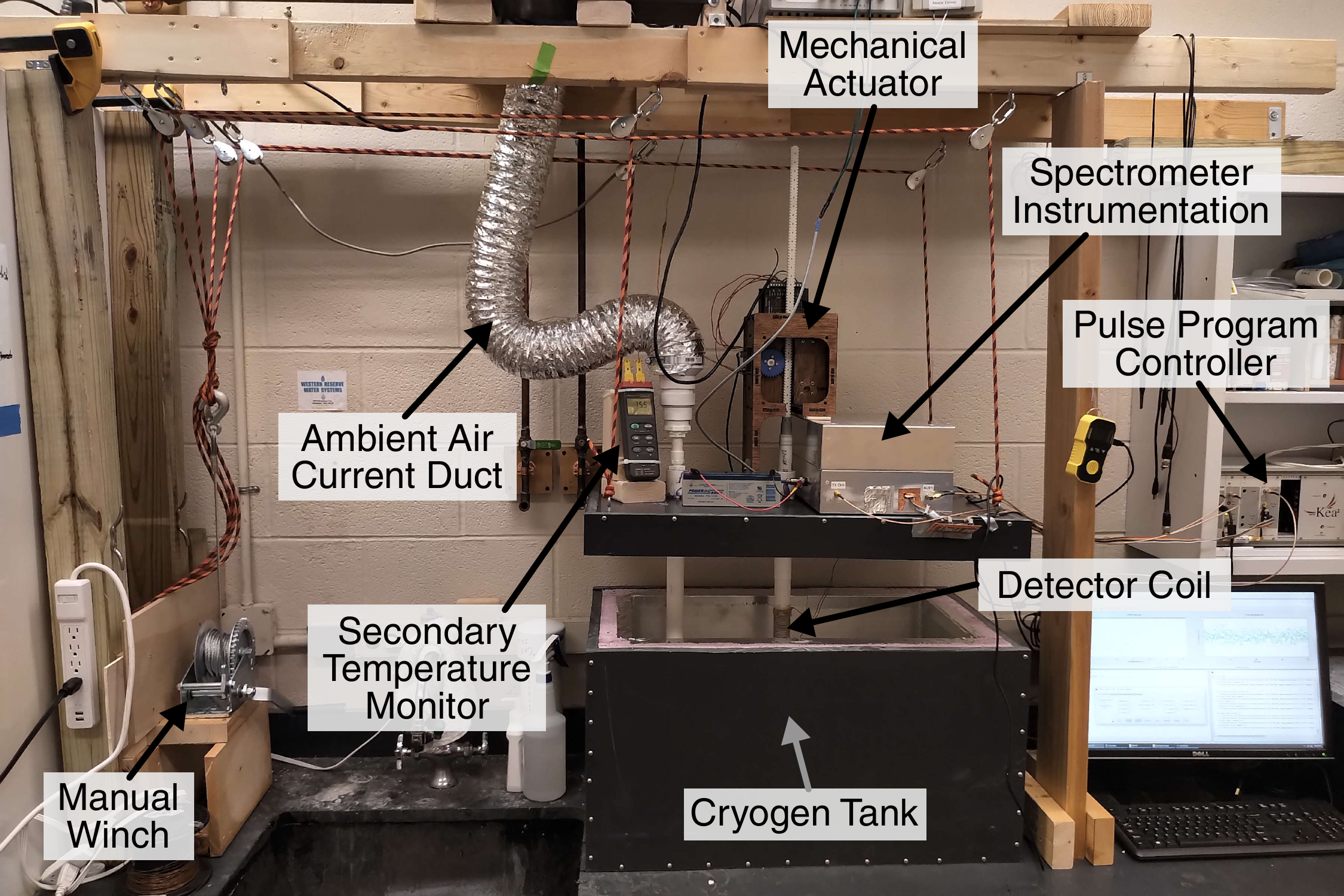}
\caption{Full implementation of the cryogenic spectrometer, including control terminal, samples, instrumentation, auxilliary power supplies, and climate controls: (top) full view, (bottom) zoomed-in view.}
\label{fig:cryo_setup_full}
\end{figure}

The complete spectrometer system is shown in Fig.~\ref{fig:cryo_setup_full} (top). It is operated from a control terminal running on a dedicated desktop computer. Due to the known long duration of experiments, the control terminal was supported by a backup battery. Backup batteries also protect the actuator and VTX power supplies and the ambient air blower's AC induction motor. Spectrometer instrumentation was also powered by battery packs to minimize noise, as described in Section~\ref{sec:power_supplies}. Thus loss of AC outlet power for up to one hour can be sustained without interruption to the experiment. Continuous data synchronization with a remote file server was implemented to ensure reliable access to data.

Spectrometer instrumentation was maintained at room temperature in a shielded steel enclosure mounted to the lid of the cryotank with a small ($\sim$2.5~mm) air gap for thermal insulation. During the experiment, the box experiences minimal cooling, though small amounts of ice have been observed to accumulate on the leads of the coil; this was removed after every $\sim$4-6 hours at cryogenic temperature using a compressed air blower at 40~psi (not shown). A particulate filter and dessicant drier were included in the air supply.

\begin{figure}
\centering
\includegraphics[width=0.7\columnwidth]{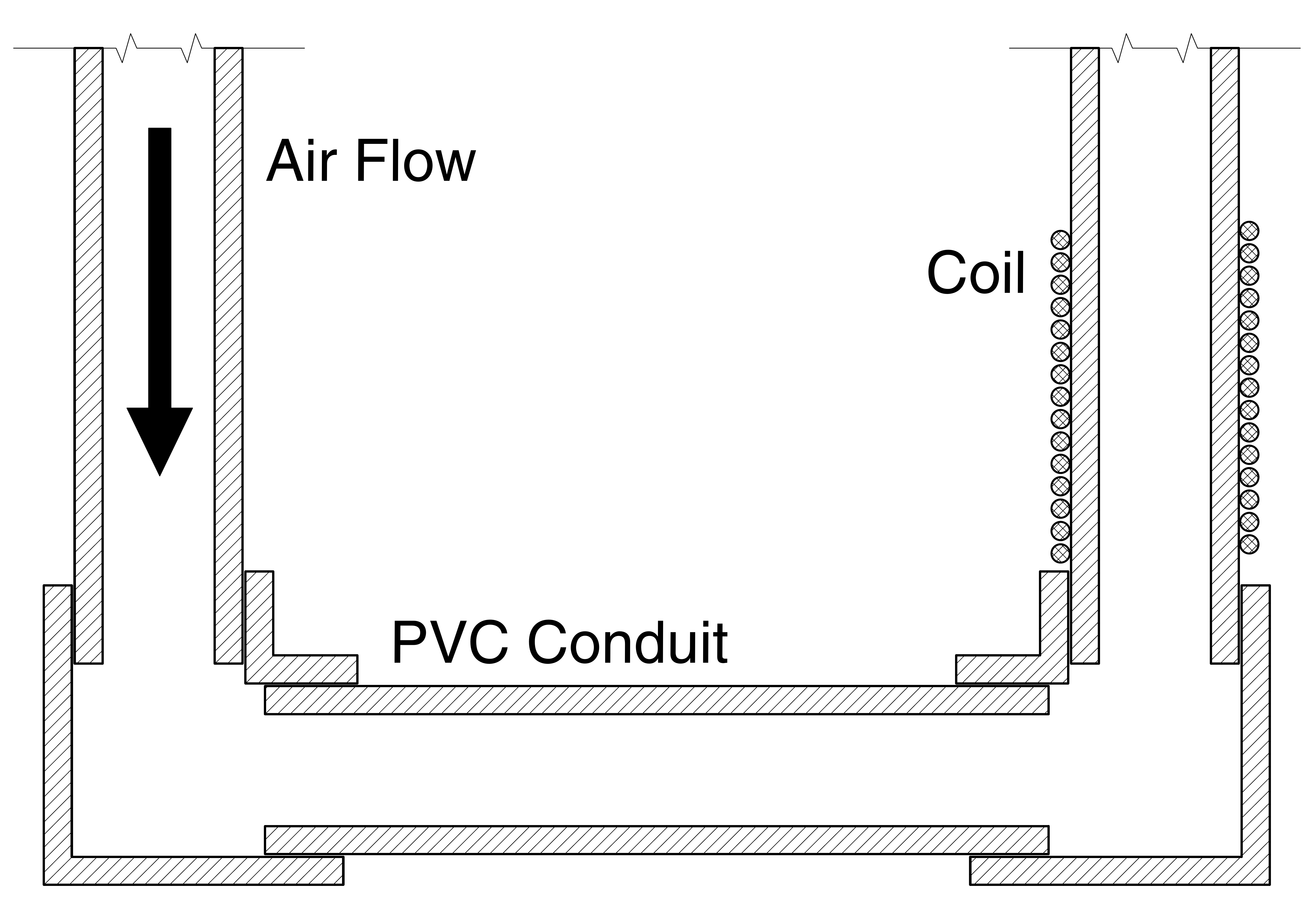}
\caption{U-bend conduit constructed by solvent welding of PVC pipes and fittings. The detector coil is wrapped around the lower section of one side.}
\label{fig:cryo_pvc_conduit}
\end{figure}

A PVC conduit formed into a U-bend was installed under the lid of the cryotank, Fig.~\ref{fig:cryo_pvc_conduit} shows the general layout. The coil was wrapped around the lower section of the right side; uniformly-spaced threads were cut in this section of pipe using a lathe before assembly of the U-bend. Both ends of the conduit are accessible from outside the cryotank to enable inspection and maintain some degree of control over temperature in the region where the sample is installed.

Additional spectrometer features are indicated in Fig.~\ref{fig:cryo_setup_full} (bottom). When fully assembled, the cryotank lid is estimated to have a mass of over 10~kg. A pulley system and manual winch is used to lift this lid in order to access the U-bend to wrap the coil and for visual inspection of the cryogen level. The secondary temperature monitor uses two K-type thermocouples to monitor the temperature of the sample and ambient space in the conduit. An ambient air blower produces a continuous air current through the U-shaped PVC pipe submerged in cryogen to reduce the temperature drop in the sample. 
A computerized linear actuator was used to apply polarization enhancement between scans. The actuator was also used to remove the sample from the cryogen between scans, thus reducing heat loss from the sample between runs to maintain consistent temperature.

\subsection{Coil Characterization in the Cryogen Tank}
\label{sec.toward_cryo_tank_v1_coil_char}

\begin{figure}
\centering
\includegraphics[width=0.8\columnwidth]{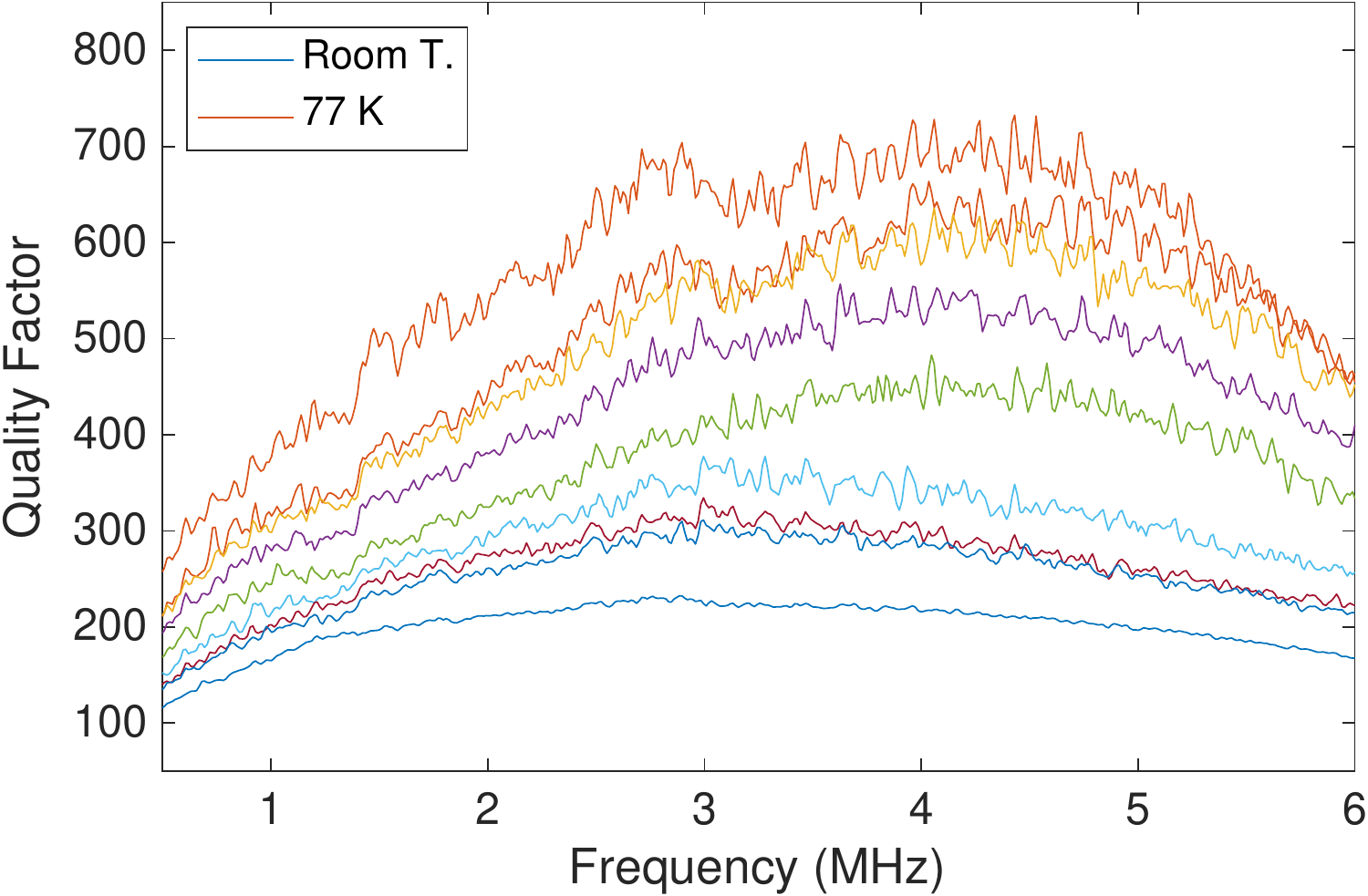} 
\caption{Measured quality factor for the test coil submerged in liquid-$N_2$. The highest curve in the plot is at 77~K. Subsequent lower curves are from data recorded at intervals as the nitrogen boiled off in the tank. The lowest curve is at room temperature. Impedance data ($R_s$ and $L_s$) was computed by de-embedding the effects of $C_p$.}
\label{fig:cryocoil_Q_HP4395A_act}
\end{figure}

The coil under test is a solenoid (design ``A'' in Table~\ref{tab:rx_coil_sampling}). It uses AWG 18 copper wire wrapped around a coil form with a 0.5~mm groove, yielding a ID of 31.75~mm, and consists of 48 turns with $\sim$1.7~mm pitch (6 turns per cm). Coil resistance and inductance was estimated from VNA data using the approach described in Section~\ref{subsubec:coil_flipper_spice}. Resistance scales linearly $\propto f$ and inductance is constant (found to be $\sim 23.5$~$\mu$H) as expected. The self resonant frequency of the coil was $f_{sr}\approx 7$~MHz. The VNA data was also used to estimate Q as a function of coil temperature $T_c$. The results, which are summarized in Fig.~\ref{fig:cryocoil_Q_HP4395A_act}, show that Q increases by approximately $3\times$ between 300~K and 77~K, in agreement with our theoretical analysis.


\subsection{Detector Tuning and Quality Factor}
Automated measurement of the probe network impedance at all possible combinations of tuning capacitor was used to produce a tuning table which reports the resonant frequency and bandwidth at all points. Impedance measurement data was also recorded for use in data visualizations. 

\begin{figure}
\centering
\includegraphics[width=0.75\columnwidth]{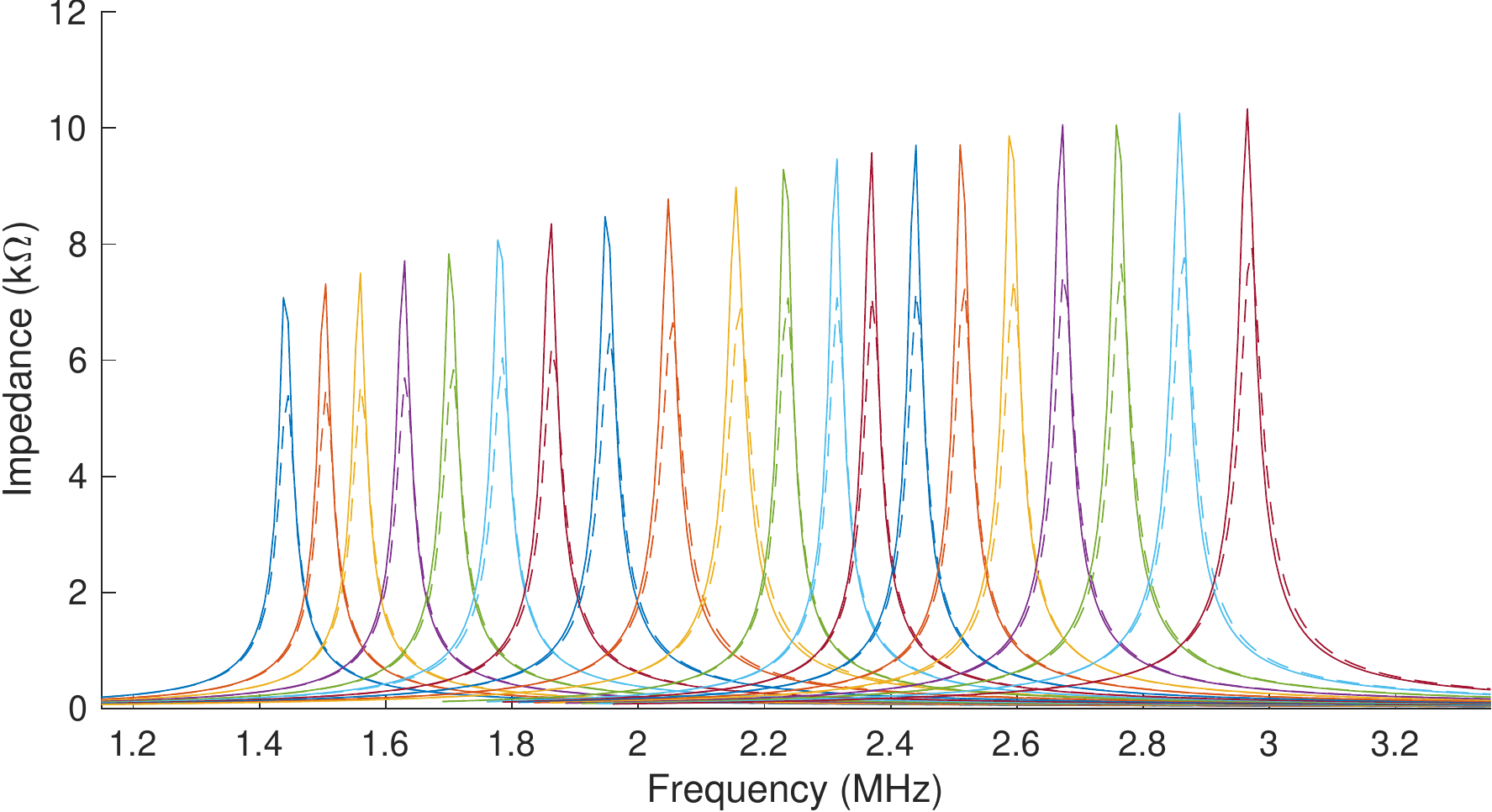}
\caption{Network impedance measurement at various tuning frequencies collected at room temperature (dashed lines) and at 77~K with the detector in parallel resonant configuration.}
\label{fig:cryo_impedance_peaks}
\end{figure}

Fig.~\ref{fig:cryo_impedance_peaks} is a plot of select measurement results across a broad range of operating frequencies. There is an unusual drop in the network quality factor around 2.9~MHz which is observed to be consistent at room temperature and 77~K. This might be the result of secondary resonant networks formed between the network analyzer and the probe as long wires are required to reach the probe in the setup. Another (more likely) possibility is that some of the capacitors in the tuning network have significantly higher ESR than others. 

\begin{figure}
\centering
\includegraphics[width=0.75\columnwidth]{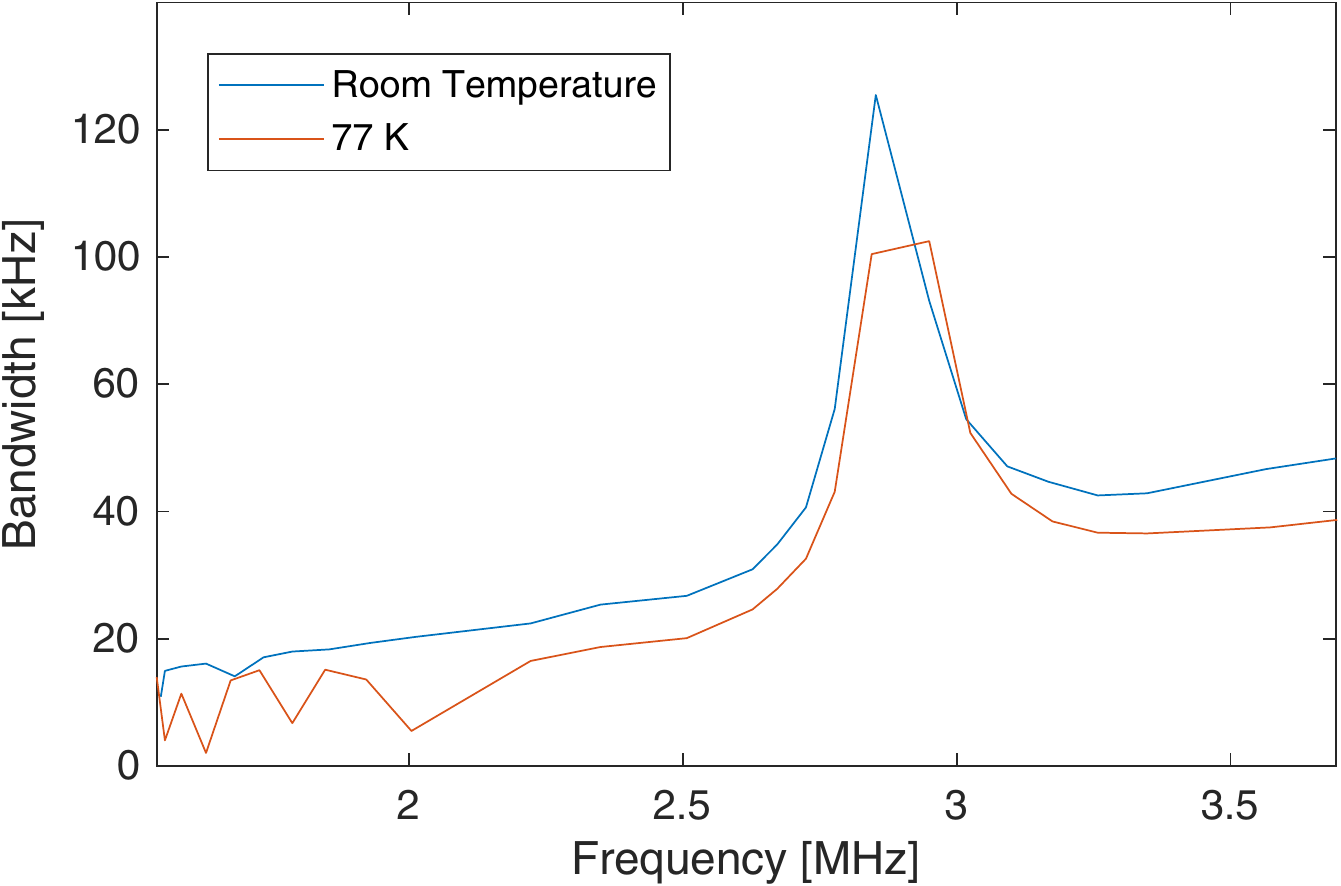}
\caption{Bandwidth computed for tuned network at resonant frequencies corresponding to the impedance data shown in Fig.~\ref{fig:cryo_impedance_peaks}.}
\label{fig:cryo_bandwidth}
\end{figure}

A resonant bandwidth measurement corresponding to selected points in the previous sweep is shown in Fig.~\ref{fig:cryo_bandwidth}. The bandwidth decreases at all points at 77~K, indicating an increase in Q as expected.

\subsection{Room Temperature Measurements}
\begin{figure}
\centering
\includegraphics[width=0.75\columnwidth]{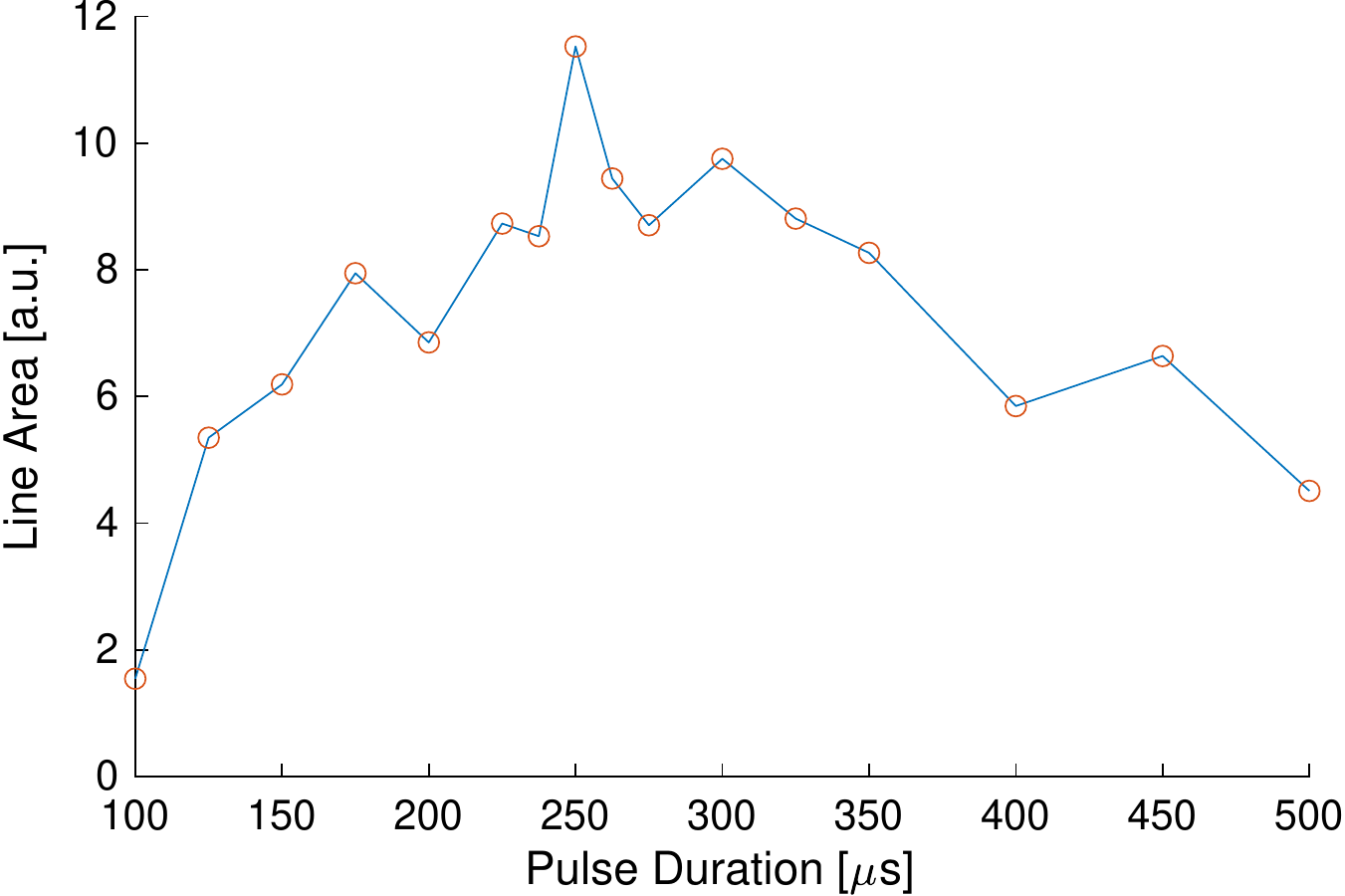}
\caption{RT nutation curve produced from pulse duration sweep at 2.564~MHz on 15~g acetaminophen (CVS brand tablets). Experimental parameters: $N=64$ scans per point, inter-experiment time of $t_{R}=20$~s, echo period $t_{E}=900$~$\mu$s, and $N_E=1000$ echoes per scan.}
\label{fig:nutation_curve}
\end{figure}

Experiments were performed using SLSE pulse sequences on a 15~g acetaminophen sample in the form of 30 CVS-brand 500~mg tablets. As the first step, a pulse duration sweep at room temperature (RT) was used to determine the RF pulse length corresponding to the optimum flip angle $\theta_{opt}$ that maximizes the echo signal amplitude. It is known that $\theta_{opt}=119.5^{\circ}$ for $I=1$ in crystalline powders~\footnote{This value assumes that the asymmetry parameter $\eta\neq 0$, which is true for the vast majority of samples}. The measured nutation curve, which is shown in Fig.~\ref{fig:nutation_curve}, also provides a valuable tool to confirm that acquired signals which appear to be echoes are in fact a result of quadrupole resonance. External environmental noise near the resonance frequency (denoted by $\omega_0$) or piezoelectric ringing may appear similar to an NQR echo, but the response of these sources does not vary with RF pulse duration.

\begin{figure}
\centering
\includegraphics[width=0.475\columnwidth]{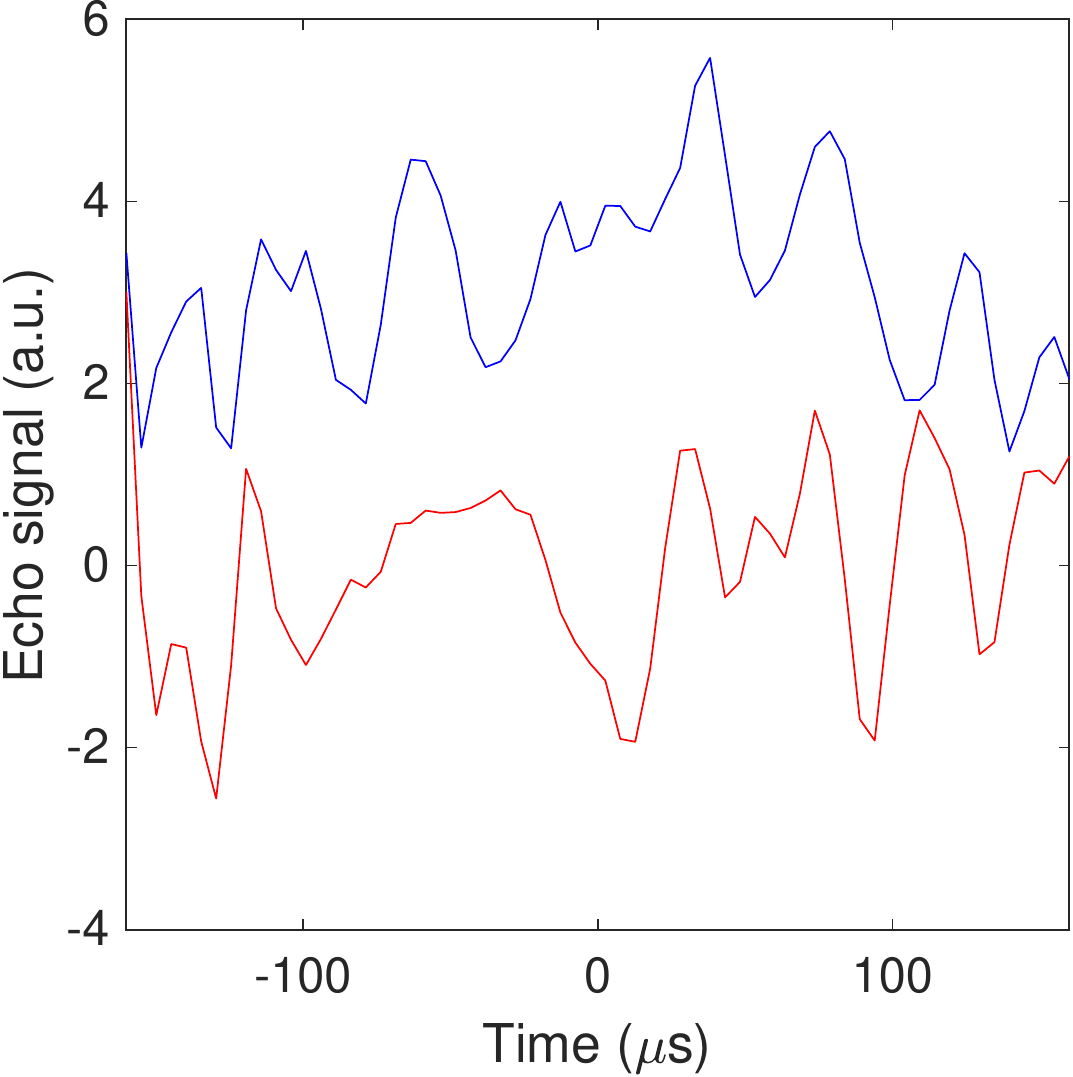}
\includegraphics[width=0.475\columnwidth]{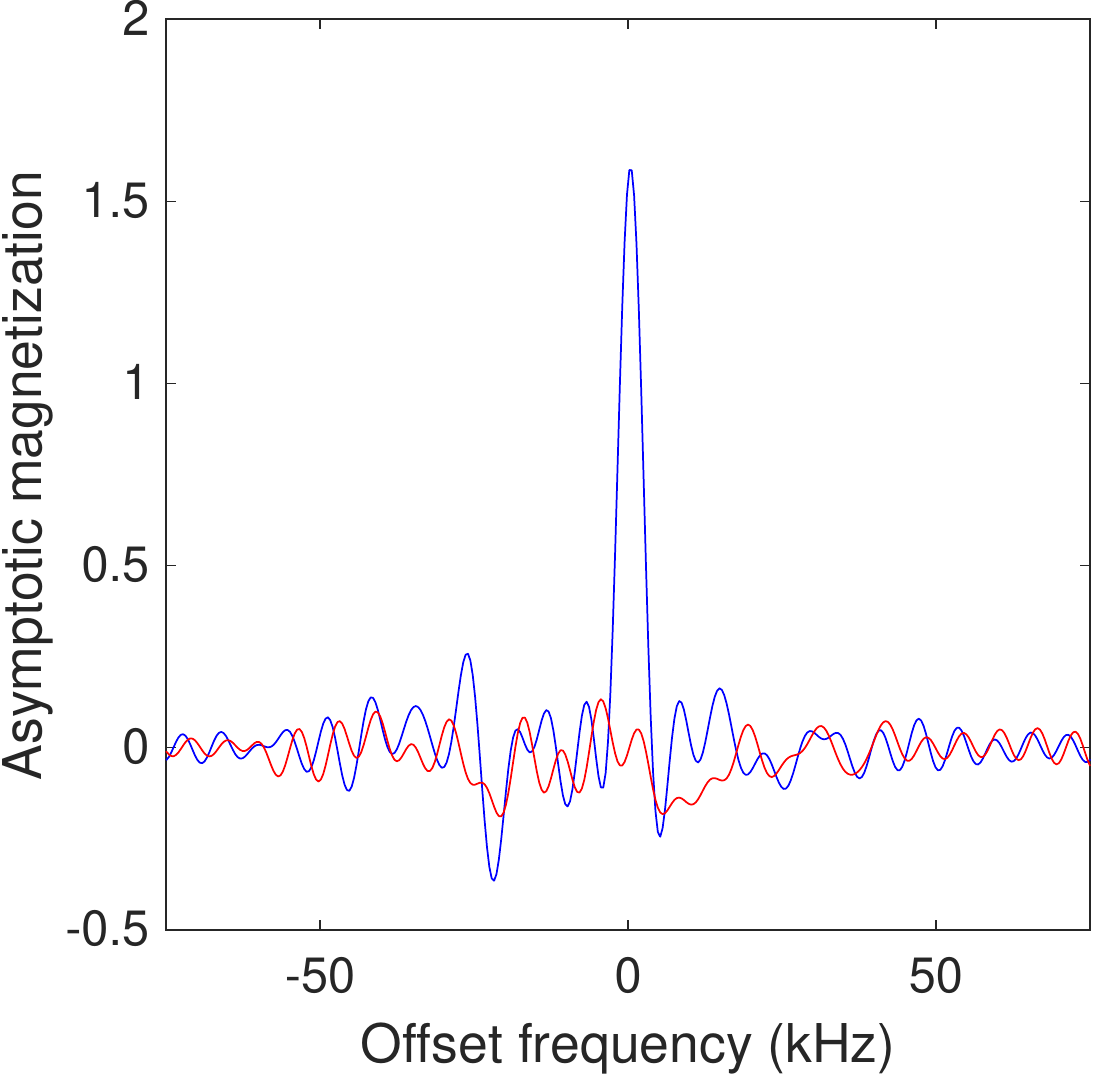}
\caption{Time domain echo (left) and corresponding spectrum (right) for a RT experiment on 15~g acetaminophen at the optimal flip angle. Experimental parameters: $N=64$ scans with pulse duration $t_{p}=250$~$\mu$s at 2.564~MHz. Blue and red traces correspond to the real and imaginary components of the data, respectively.}
\label{fig:nutation_curve_250_peak}
\end{figure}

The measured time domain echo sum and the corresponding spectrum for measurements of the $\nu_+$ transition at the optimal flip angle are shown in Fig.~\ref{fig:nutation_curve_250_peak}. Note that the complex acquired data was phase rotated to ensure that all the signal energy lies in the real channel; such phase sensitive detection improves $SNR_e$ by a factor of 2.

%
%

%


\subsection{Low Temperature Measurements}
NQR resonant frequencies are known to be temperature-dependent. Acetaminophen is an unusual sample as it is one of the few compounds for which the documented frequency shift for $^{14}$N NQR has a positive temperature coefficient. Barras et al. have measured  a positive temperature coefficient for the $\nu_+$ line of approximately $+69$~Hz/K at temperatures between 280 and 310~K~\cite{nqr:kyriakidou2015_batch_descrimination}. However there is no reference data on this compound at lower temperatures. Using the known trend around room temperature~\footnote{Supplemental data in~\cite{nqr:kyriakidou2015_batch_descrimination} suggest the relation $\omega(T)=2562 + 0.069$ where $\omega$ is in kHz.} we expect $\nu_+$ would decrease to 2.547~MHz at 77~K assuming linearity. However, prior measurements of other compounds shows that the magnitude of the temperature coefficient decreases below $\sim$120~K~\cite{nqr:grechishkin1988, nqr:osokin1980}, so this result is expected to be an under-estimate. 

SLSE experiments were repeated in dry ice (194.7~K) and liquid-N$_{2}$ (77~K); both were chosen for their low cost and wide availability. Preventing heat flux in the sample was found to be difficult due to the large temperature gradient between ambient air and the cryogen, particularly as ice collects gradually while an experiment runs. Initially the sample was inserted into the detector for 2~s and removed between scans during the 20~s repetition time. However, the sample temperature was noted to drop an average of 20$^{\circ}$ from ambient and vary by as much as 10$^{\circ}$ during the experiment, which is enough for $\omega_0$ to vary by 1-2~kHz during acquisition. Thus, we allow the sample to cool to thermal equilibrium inside the U-bend and perform a frequency sweep to search for the new resonant frequency. As discussed in Section~\ref{subsec:temp_sens}, the reduced sample temperature further increases SNR by increasing the NQR signal amplitude. The T$_1$ relaxation time is also expected to increase at lower temperatures (approximately $\propto E_{a}/(RT_s)$), so the inter-experiment time was increased from 20~s to 60~s.

Measurements in dry ice (194.7~K) used a frequency sweep with a step size of 1~kHz. The results reveal a resonant frequency of 2.552~MHz, which is slightly lower than expected from a linear extrapolation of near-RT data. Additionally, the SNR per scan increased by $2.47\times$ compared to RT, which is about $2\times$ lower than the expected increase of $4.64\times$. This discrepancy is probably due to the additional resistance in the resonant network, which consists of the ESR of the tuning capacitors, $R_{DS,ON}$ of the active duplexer switches, contact resistance of the relays, and resistance of the PCB traces and connecting wires. These components remain at RT, so their noise does not improve with reduced probe temperature.

\begin{figure}
\centering
\includegraphics[width=0.475\columnwidth]{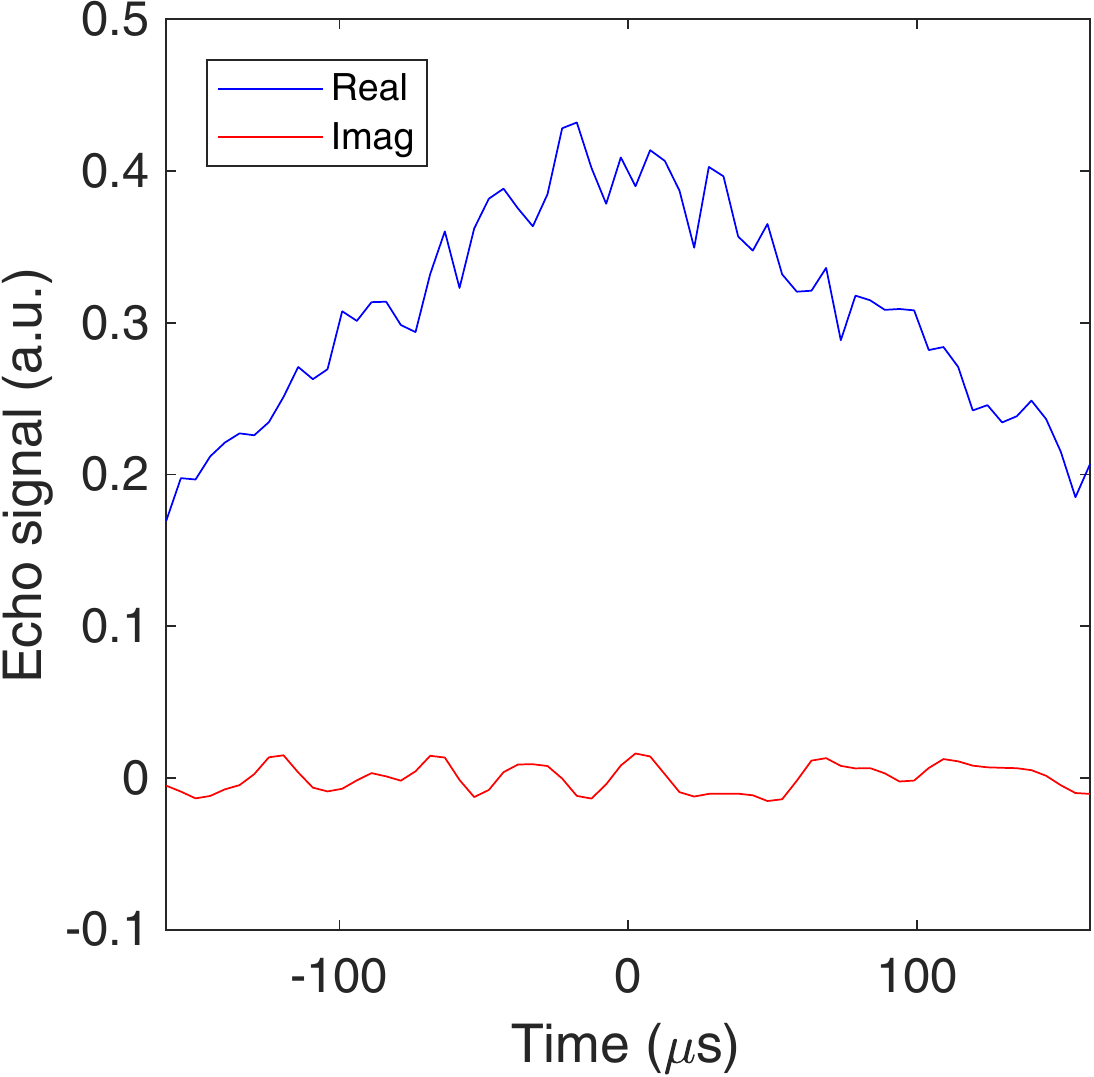}
\includegraphics[width=0.475\columnwidth]{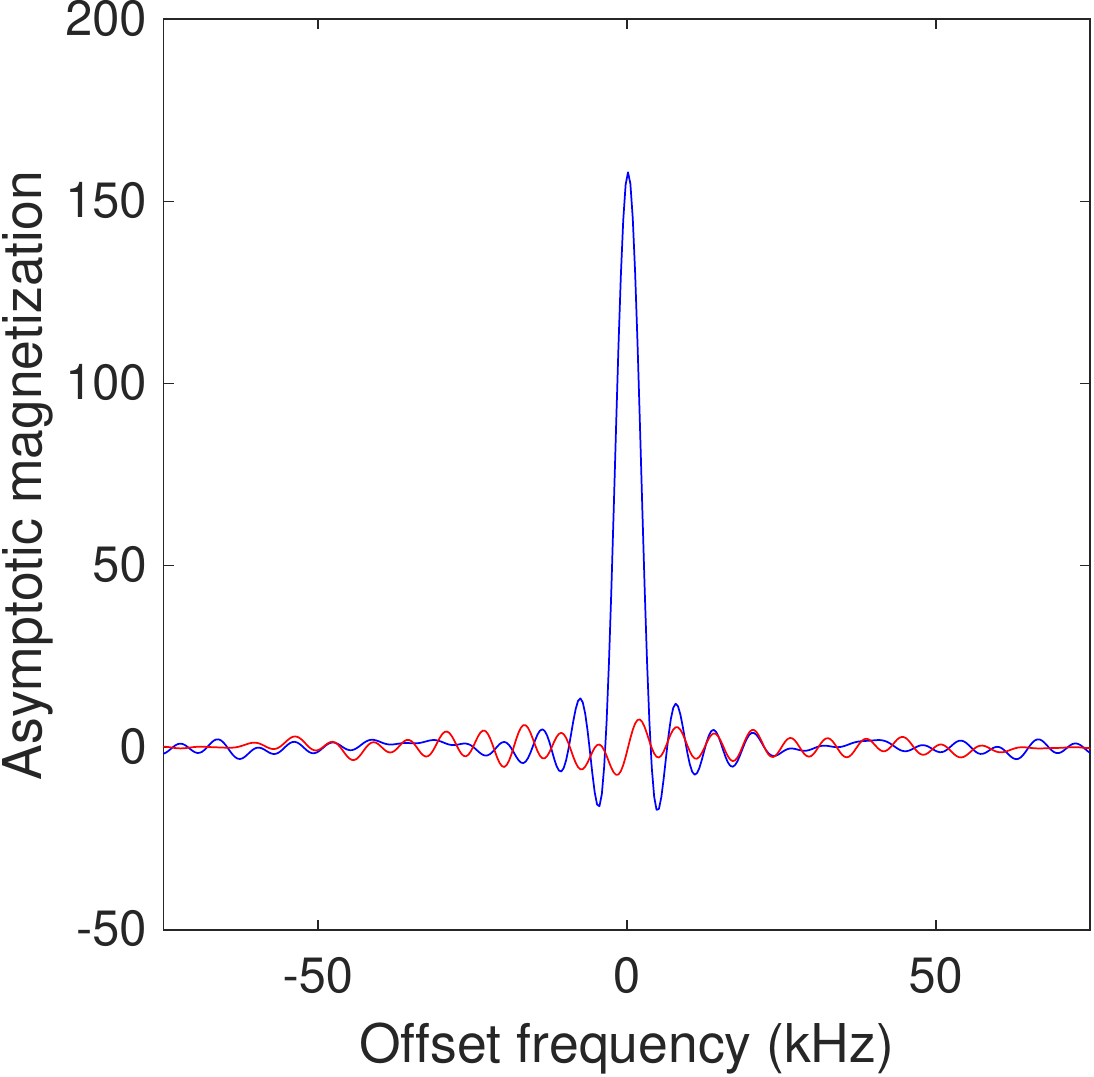}
\caption{Time domain echo (left) and corresponding spectrum (right) for an experiment on 15~g acetaminophen with probe and sample at 77~K and the optimal flip angle. Experimental parameters: $N=8$ scans, RF frequency $=2.542$~MHz. Blue and red traces correspond to the real and imaginary components of the data, respectively.}
\label{fig:ln2_peak}
\end{figure}

Measurements in liquid-N$_{2}$ reveal a resonant frequency of 2.542~MHz, which in general agreement with our prediction based on the near-RT and dry ice data. The measured time- and frequency-domain data after $N=8$ scans is shown in Fig.~\ref{fig:ln2_peak}. The expected total SNR enhancement factor at 77~K is 160$\times$ (when the sample is cooled in addition to the coil). The experimentally realized enhancement factor is 88.4$\times$, which is again about $2\times$ lower than predicted and likely limited by the additional RT resistance in the resonant network.

\begin{figure}
\centering
\includegraphics[width=0.8\columnwidth]{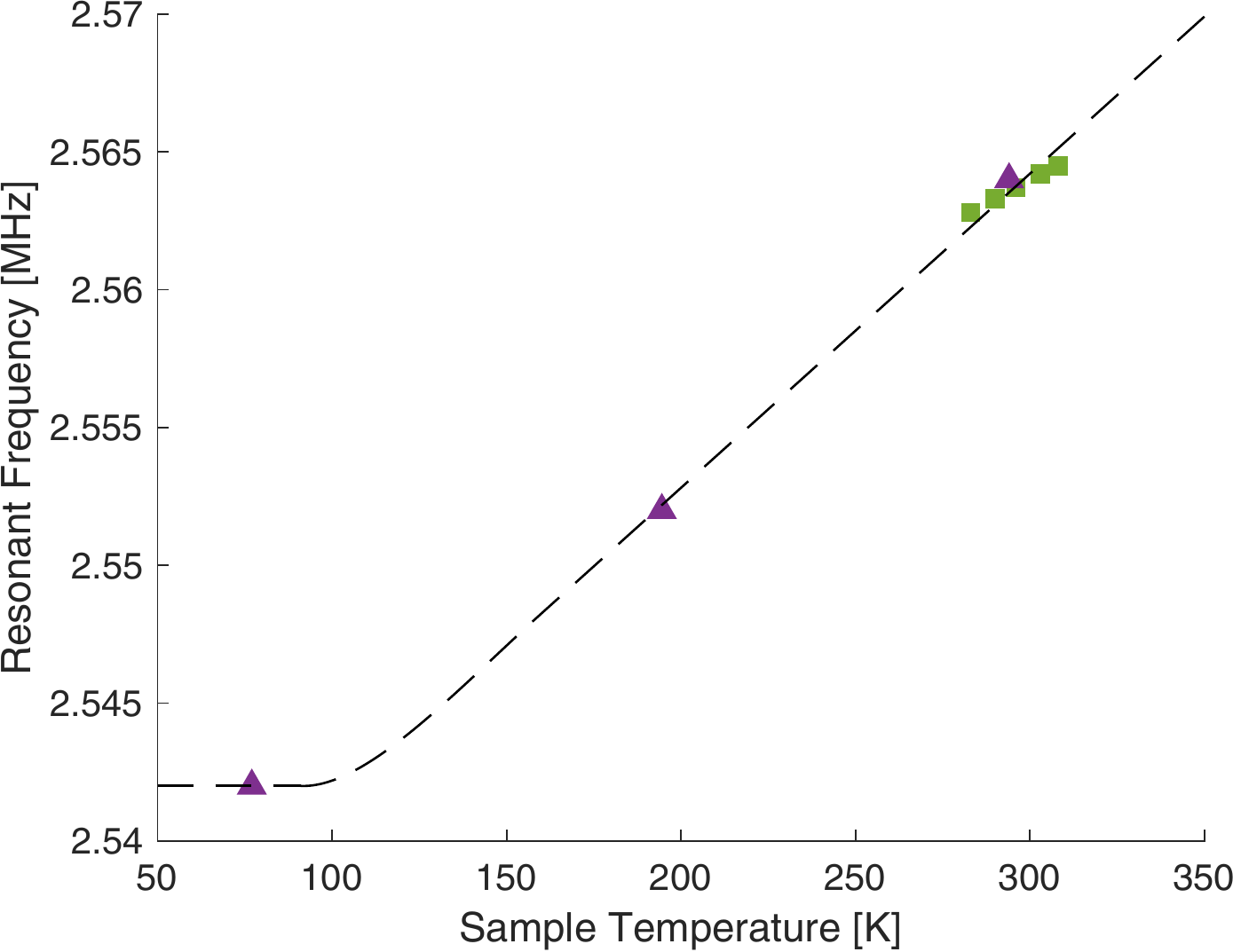}
\caption{Summary of measured resonance frequency of the $\nu_+$ transition of acetaminophen versus sample temperature. Green squares denote prior data from Barras et al.~\cite{nqr:kyriakidou2015_batch_descrimination}, while the other points were measured using our proposed spectrometer. The dashed line shows a numerical fit to the data.}
\label{fig:temperature_trend}
\end{figure}

Fig.~\ref{fig:temperature_trend} summarizes the observed resonance frequency of the $\nu_+$ transition of acetaminophen versus sample temperature. While the number of data points below RT is limited, the data suggests that the temperature coefficient decreases below $\sim$120~K, as seen in prior work on other compounds~\cite{nqr:grechishkin1988, nqr:osokin1980}.

\subsection{Further Enhancement using PE-NQR}
In previous work, polarization enhancement (PE) by adiabatic population transfer from nearby protons has been demonstrated to improve $SNR_e$ of $^{14}$N NQR by 25-64$\times$~\cite{glickstein-automated-PENQR}. Unlike the cryogenic approach, signal enhancement by PE-NQR is not a universally applicable technique; it is not effective if there are no protons close to the target $^{14}$N nucleus. In addition, the prepolarization time required varies greatly between samples and is dependent on the $^1$H NMR spin relaxation time $T_{1}$, which varies widely between compounds. Furthermore, the enhancement factor is proportional to $\nu_{H}/\nu_{N}$, meaning that is is less effective for high frequency lines. Thus we need to know some information about the compound under study in order to set the pre-polarization time, predict the effective SNR, and determine the required number of scans. Nevertheless the technique has great value in some cases, especially for low frequency ($<2$~MHz) lines.

Here we briefly describe the potential combination of PE with cryogenic noise reduction for even higher SNR per scan. A probe designed for this purpose should allow the sample to be first pre-polarized in a static field $B_0$ and then adiabatically demagnetized to zero-field prior to the NQR measurement. Since electromagnets generate significant heat (making them incompatible with cryogenic operation), we instead generate the required time-varying $B_{0}$ field by using a computerized mechanical actuator to move the sample outside a permanent magnet, as in our earlier work~\cite{glickstein-automated-PENQR}. 

For simplicity, consider a mechanically simple design in which both the actuator and magnet are immersed within the cryogenic liquid. While neodymium alloy permanent magnets exhibit brittle behavior at all temperatures down to 4~K, their mechanical properties (including elastic modulus $E$ and breaking stress $\sigma$) improve at low temperatures~\cite{magnets:vedrine1990}. Furthermore, the remanent field strength $B_{r}$ increases with decreasing temperature. However, NdFeB material undergoes spin reorientation at 135~K, which can decrease the magnetic flux by up to 14\% below this temperature. Fortunately, this effect does not cripple the effectively increased field strength of NdFeB magnets when cooled to 77~K. To demonstrate this point, we measured the field strength of a custom NdFeB Halbach dipole array magnet large enough to accommodate the sample after immersion in liquid-N$_{2}$ within the cryotank. The field strength increased by 10\% (from 706~mT to 777~mT) between RT and 77~K. The corresponding signal enhancement factor (EF) for the $\nu_+$ transition of acetaminophen is predicted to be $16.5\times$~\cite{chen2015authentication}. Thus, combining PE-NQR with cryogenic operation (at 77~K) is expected to increase SNR per scan by a factor of $88.4\times (16.5)^2\approx 2.4\times 10^{4}$. The result would be approximately 2 orders of magnitude improvement in sensitivity (i.e., limit of detection) and/or 4 orders of magnitude reduction in averaging time required to obtain a given SNR. Similar improvements are expected for $^{14}$N sites in other molecules.

\section{Conclusion}
\label{sec:conclusion}

We have described a custom benchtop spectrometer for $^{14}$N NQR measurements. The system allows both the detector and sample temperature to be reduced to 77~K (and potentially even lower) to greatly improve the signal-to-noise ratio (SNR) of NQR scans, thus increasing sensitivity and lowering the limit of detection for a given experiment duration. Low-loss switches allow the NQR sample probe to be reconfigured from series-tuned in transmit mode (thus eliminating HV power supplies) to parallel-tuned in receive mode (thus minimizing receiver RF). The probe also uses a dual-array tuning network to enable its resonant frequency to be programmed over a broad frequency range, while a noiseless feedback damping circuit ensures adequate bandwidth in receive-mode. The spectrometer implements several self-monitoring features to enable autonomous and/or remotely-controlled operation. Tests on an acetaminophen sample confirm the expected increase in SNR per scan at cryogenic temperatures.

\begin{acknowledgments}
The authors wish to thank Dr. Michael Malone for several useful discussions.
\end{acknowledgments}

\section*{Data Availability}
The data that support the findings of this study are available from the corresponding author
upon reasonable request.

\bibliography{references}

\end{document}